\documentclass[aps,pra,superscriptaddress,epsfigure,twocolumn,nofootinbib]{revtex4-1}
\usepackage{bbm} 
\usepackage{graphicx}
\usepackage[dvipsnames]{xcolor}
\usepackage{amsmath,amssymb}    
\usepackage{latexsym}
\usepackage{amsfonts}   
\usepackage{amssymb}
\usepackage{hyperref}
\usepackage{color}
\usepackage{physics}
\newcommand{\ii}{\mathbbm{i}}
\begin{document}

\title{ {Driven quantum harmonic oscillators: A working medium for thermal machines}}
\author{Heather Leitch}
\affiliation{Centre  for  Theoretical  Atomic,  Molecular  and  Optical  Physics, Queen's  University  Belfast,  Belfast  BT7 1NN,  United  Kingdom}
\author{Nicol\`o Piccione}
\author{Bruno Bellomo}
\affiliation{Institut UTINAM, CNRS UMR 6213, Universit\'{e} Bourgogne Franche-Comt\'{e}, Observatoire des Sciences de l'Univers THETA, 41 bis avenue de l'Observatoire, F-25010 Besan\c{c}on, France}
\author{Gabriele De Chiara}
\affiliation{Centre  for  Theoretical  Atomic,  Molecular  and  Optical  Physics, Queen's  University  Belfast,  Belfast  BT7 1NN,  United  Kingdom}

\begin{abstract}
The study of quantum thermodynamics is key to the development of quantum thermal machines.  {In contrast to most of the previous proposals based on discrete strokes, here we consider a working substance that is permanently coupled to two or more baths at different temperatures and continuously driven. To this end, we investigate  parametrically driven quantum harmonic oscillators coupled to heat baths via a collision model.} Using a thermodynamically consistent local master equation, we derive the heat flows and  {power} of the working device which can operate as an engine, refrigerator or accelerator and analyze the instantaneous and average efficiencies and coefficients of performance. Studying the regimes of both slow and fast driving of the system, we find that an increased driving frequency can lead to a change of functioning to a dissipator. Finally, we investigate the effect of squeezing one of the thermal baths: it leads to an apparent higher efficiency compared to the corresponding Carnot value of an equilibrium bath with the same temperature and to sustained entanglement between the working substance oscillators in the limit cycle.
\end{abstract}

\maketitle

\section{Introduction}
The study of quantum thermal machines and their potential advantages over their classical counterparts has gained much interest in recent years~ {\cite{KosloffARPC2014,Deffner19,Mitchison19}, after the first experimental realizations had been demonstrated \cite{rossnagel2016single,MaslennikovNatComm19,PetersonPRL2019,GluzaPRXQ2021}}.
The theory of open quantum systems provides us with the tools to calculate the dynamics of the working substance interacting with one or more much larger environments. 
Studying the energy flows which result from a system interacting with heat baths of different temperatures is key to understanding the conditions required for thermal machines such as engines and refrigerators to operate.

The working substance itself may consist of  quantum systems of different nature, including quantum harmonic oscillators (QHO's)  {\cite{KosloffEntropy2017,SinghPRE2020,ReidEPL2017,DeffnerEntropy2018,SerafiniPRL2020}}, qubits  {\cite{QuanPRE2007,HewgillPRA2018,SolfanelliPRB2020}} or multilevel systems  {\cite{QuanPRE2005,campisi2016power,NiedenzuNJP2018,PiccionePhysRevA2021}}.
Thermal machines can be normally split into three categories depending on how the working substance is operated upon: autonomous (where the Hamiltonian of the working substance is time-independent), discrete-stroke (e.g. the Otto or Carnot cycle, in which the evolution is split into finite-time intervals in which the baths are connected and disconnected) and continuously driven (in which the system is permanently coupled to the baths and its Hamiltonian is time-dependent). Though the equivalence of these categories has been proven \cite{UzdinPRX2015}, for practical applications it is still significant to maintain the separation. Besides, a no-go theorem exists for autonomous linearly-coupled QHO to operate as an absorption refrigerator \cite{MartinezPRL2013}.

In this paper, we  consider thermal machines of the third category: a network of driven QHO's permanently coupled to heat baths. The evolution of the system, in the limit of weak coupling, is described by an adiabatic Markovian Lindblad master equation within the local approximation. The thermodynamic consistency of local master equations is guaranteed by their microscopic model based on the so-called repeated interactions or collision models,  Refs.~\cite{DeChiaraNJP2018} and \cite{HewgillPRR2021} (see Refs.\cite{CampbellEPL2021} and \cite{ciccarello2021quantum} for recent reviews).  

We focus on working substances made of one and two driven QHO's interacting with two baths and find the conditions required for the system to operate as an engine, a refrigerator or an accelerator. The effect of both slow and fast driving is explored and we see that the increased driving speed leads to a larger injection of  {power} into the system  {causing heat to flow from the system into both baths. This operation is referred to as a dissipator or heater \cite{SolfanelliPRB2020}.}
We study the efficiency and the coefficient of performance (COP) of the device when operating as an engine and as a refrigerator, respectively. We show that driving the system oscillators causes the  {power} to oscillate periodically in time. In some cases, this leads to the instantaneous coefficient of performance of a refrigerator being higher than the Otto value.
We also show how the presence of squeezing in one of the baths effectively raises the temperature of the bath resulting in an effective higher Carnot efficiency. Squeezing is also required for the generation of entanglement in a working medium made of two QHO's.

This paper is organized as follows. In Sec.~\ref{sec:model} we explain our microscopic model and derive the relevant thermodynamic quantities. In Sec.~\ref{sec:cov1} we illustrate the results for the simple case of a single QHO coupled to two baths at different temperatures while in Sec.~\ref{sec:twoosc} we extend our study to a working substance of two QHO's. In Sec.~\ref{sec:squeezed} we introduce squeezing in the hotter bath in the two-oscillator model showing an (apparent) enhanced efficiency and the sustenance of entanglement between the two oscillators in the long-time limit.
Finally, in Sec.~\ref{sec:conclusions} we summarize our findings and conclude.
 Some details of our analysis can be found in the appendices.

\section{The Model}
\label{sec:model}

\begin{figure}[t]
\begin{centering}
\includegraphics[width=1\columnwidth]{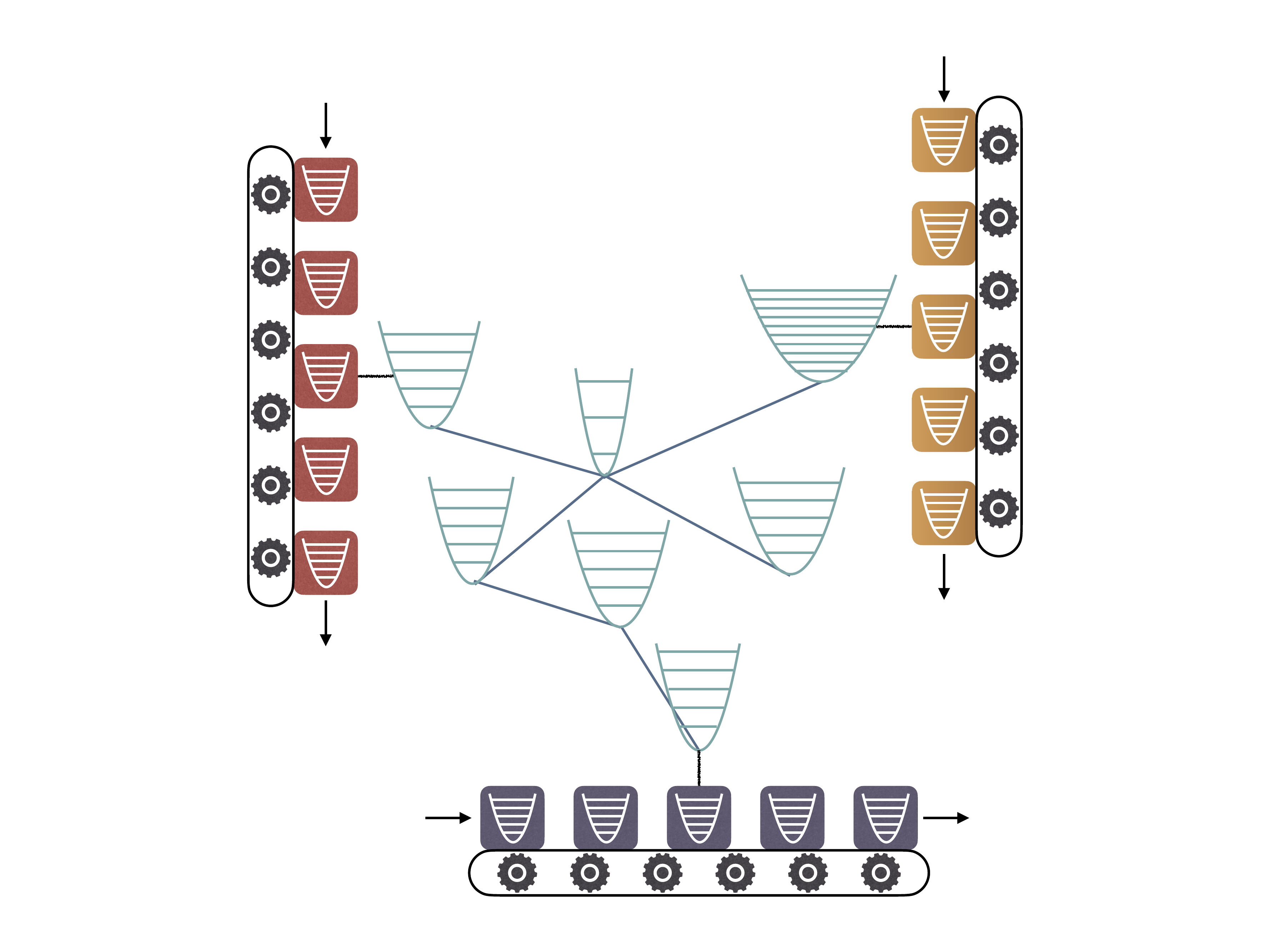}
\caption{A network of $N$ coupled quantum harmonic oscillators interacting with $N_B$ heat baths via the collision model.}
\label{fig:diagram}
\end{centering}
\end{figure}

In this section we describe the general setup we are going to consider in this paper. The system is composed of $N$ driven quantum harmonic oscillators (QHO's) coupled to $N_B$ baths via a collision model, as depicted in Fig.~\ref{fig:diagram}. 
Let us define the position $x_i$ and momentum $p_i$ operators of the ith oscillator in the system subject to the time-dependent frequency $\omega_i(t)$ and with unit mass.
The time-dependent Hamiltonian for the system is (setting $\hbar=1$),
\begin{eqnarray}
H_S(t) &=&\frac 12 \sum_{i = 1}^{N}\left[p^2_i+ \omega^2_i(t) x_i^2 \right]   + 
\nonumber\\
&+&\sum_{\substack{ {i<j}}} \lambda_{ij}\left[ a_i^\dagger(t) a_{j}(t) + a_i(t) a_{j}^\dagger(t) \right],
\label{eq:HS}
\end{eqnarray}
where the time-dependent annihilation operators $a_i(t)$ are defined as
\begin{equation}
a_i(t)=\sqrt{\frac{\omega_i(t)}{2}}x_i+\ii\sqrt{\frac{1}{2\omega_i(t)}}p_i.
\end{equation}
In Eq.~\eqref{eq:HS}, $\lambda_{ij}$ (assumed to be real) is the strength of the beam-splitter-type coupling between the ith and jth system oscillator and is equal to zero if there is no coupling between them. The number-conserving form of the coupling term ensures that excitations can hop between coupled oscillators but are not destroyed or created in the process.  {The presence of counter-rotating terms in the coupling, arising for instance from a position-position coupling, tends to hinder the transport efficiency as discussed in Ref.~\cite{DeChiaraNJP2018}. The beam-splitter-type coupling of Eq.~\eqref{eq:HS} may be implemented in photonic waveguides as discussed in  Ref.~\cite{ReidEPL2017}.}

The oscillators' frequencies are modulated in time according to
\begin{equation}
\label{eq:omegat}
\omega_i(t) = \omega_{0,i}  + \delta \omega_i \sin(\theta_i t),
\end{equation}
with $\delta\omega_i<\omega_{0,i}$ to ensure positivity of the frequencies and where $\theta_i$ is the modulation frequency.

The baths are modeled as an infinite ensemble of QHO's such that the total bath Hamiltonian is $H_B =\sum_{\alpha=1}^{N_B} H_{B,\alpha}$ where each individual bath Hamiltonian reads
\begin{equation}
H_{B,\alpha}= \frac{1}{2}\sum_{j=1}^\infty \left(P^2_{\alpha,j}+\Omega_\alpha^2 X^2_{\alpha,j}  \right),
\end{equation}
where we have defined the corresponding position $X_{\alpha,j}$ and momentum $P_{\alpha,j}$  {operators} of the jth oscillator in the $\alpha$th bath. We also define the corresponding time-independent annihilation operators
\begin{equation}
b_{\alpha,j}=\sqrt{\frac{\Omega_\alpha}{2}}X_{\alpha,j}+\ii\sqrt{\frac{1}{2\Omega_\alpha}}P_{\alpha,j}.
\end{equation}

Note that we are considering the case where all the oscillators in each bath have the same frequency. We assume that the oscillators in the baths are initially prepared in a thermal state at their respective temperature $1/\beta_\alpha$ (setting $k_B = 1$),
\begin{equation}\label{eqthermal}
\rho_{B,\alpha} = \frac{e^{-\beta_\alpha H_{B,\alpha}}}{Z_\alpha} ,\;\;\;\;\;\;\;\;\;\;\;\;\;\;\;
Z_\alpha = \text{Tr}\left[ e^{-\beta_\alpha H_{B,\alpha}} \right],
\end{equation}
and interact with the system by means of repeated interactions or collisions. 
At each time step, one oscillator from each bath interacts with one QHO in the system for a short time period $\tau$. Each bath interacts always with the same QHO. This interaction is described by the Hamiltonian
\begin{equation}
H_I(t) =\sum_{i=1}^N\sum_{\alpha=1}^{N_B} \sum_{j=1}^\infty \frac{g_{i,\alpha} \Theta_j(t)}{\sqrt{\tau}}\left[ a_{i}^\dagger(t) b_{\alpha,j} + a_{i}(t) b_{\alpha,j}^\dagger \right],
\label{eq:HI}
\end{equation}
where $g_{i,\alpha}$ is the coupling between the ith system oscillator and the bath $\alpha$ and
where 
\begin{equation}
\Theta_j (t) =
\begin{cases}
1 & \text{for} \; t_j \leq t \leq t_j + \tau,\\
0 & \text{otherwise},
\end{cases}
\end{equation}
with $t_j$ being the time at which the jth bath oscillator starts interacting with the system. We assume $t_1=0$ and that there is no time between subsequent collisions. 
In summary, the Hamiltonian of the complete system and baths is
\begin{equation}
H(t) = H_S(t) + H_B + H_I(t).    
\end{equation}
A technical note: the $1/\sqrt{\tau}$ scaling in Eq.~\ref{eq:HI} ensures the consistency in the continuous limit ($\tau\to 0$) we are now going to describe and naturally arises in quantum optical systems~\cite{ciccarello2021quantum}.

We now introduce the adiabatic master equation for the evolution of the system's density matrix in which non-adiabatic terms arising in the dissipative part of the evolution are absent. To this end we assume that the collision time $\tau$ is the shortest timescale in the problem and that the modulation frequency $\theta_i$ corresponds to a much slower timescale~\cite{AlbashNJP2012}. This assumption results in a master equation that does not contain derivatives of the system modulated frequencies explicitly.  {The derivation then follows closely the one presented in Ref.~\cite{DeChiaraNJP2018}, where an expansion up to order $\tau^2$ for the evolved density matrix is considered. 
Making the limit $\tau \to 0$ leads to the adiabatic Markovian master equation},
\begin{eqnarray}
\dot{\rho}_S  &=& - \ii \comm{H_S(t)}{\rho_S} 
\nonumber \\
& +& \sum_{i=1}^N \sum_{\alpha=1}^{N_B} g_{i,\alpha}^2 n_\alpha \left[a_{i}^\dagger \rho_Sa_{i} -\frac{1}{2}\left\{a_{i}a_{i}^\dagger ,\rho_S\right\}\right]
\label{eq:me} \nonumber
\\
& +&\sum_{i=1}^N  \sum_{\alpha=1}^{N_B} g_{i,\alpha}^2 (n_\alpha+1) \left[a_{i} \rho_Sa_{i}^\dagger -\frac{1}{2}\left\{a_{i}^\dagger a_{i} ,\rho_S\right\}\right], \nonumber\\
\end{eqnarray}
where for brevity we have omitted the time-dependence of $\rho_S(t)$ and $a_{i}(t)$. We have also defined the thermal occupation of each bath oscillator as
\begin{equation}
n_\alpha = \frac{1}{e^{\beta_\alpha \Omega_\alpha}-1}.
\end{equation}
 {Note that Eq.~\eqref{eq:me} is a local master equation (whose validity is well justified irrespective of the magnitude of the interaction between system oscillators, see Ref.~\cite{DeChiaraNJP2018}) which, for a suitable choice of the parameters, exhibits the presence of exceptional points \cite{Scali2021localmaster}.}

Following Ref.~\cite{DeChiaraNJP2018}, we obtain the heat current to the bath $\alpha$,
\begin{eqnarray}
\dot Q_\alpha (t) &=& \lim_{\tau \rightarrow 0} \frac{1}{\tau} \text{Tr}\left[ H_{B,\alpha} \left(\rho_{B,\alpha}(t+\tau) -  \rho_{B,\alpha}(t)\right)\right]
\nonumber\\
&=&\sum_{i=1}^N g^2_{i,\alpha} \Omega_\alpha\left[n_\alpha-\langle a_{i}^\dagger a_{i}\rangle\right ],
\label{eq:Qdot}
\end{eqnarray}
and the  {power} done on the system,
\begin{eqnarray}
\dot W(t)  &=& \lim_{\tau \rightarrow 0} \frac{1}{\tau} \text{Tr}\left[ H_S (t+\tau) \rho_S(t+\tau) -  H_S(t) \rho_S(t)\right] 
\nonumber
\\
&-& \sum_{\alpha} \dot{Q}_\alpha(t),
\end{eqnarray}
giving
\begin{eqnarray}
\dot W (t)&=& \sum_{i=1}^N \sum_{\alpha=1}^{N_B} g^2_{i,\alpha} [\omega_{i}(t)-\Omega_\alpha]
\left[n_\alpha-\langle a_{i}^\dagger a_{i}\rangle\right ]+
\nonumber \\
&+& \sum_{i=1}^N \sum_{\alpha=1}^{N_B} \dot{\omega}_{i}(t)\frac{2 \expval{a_{i}^\dagger a_{i}} +1}{2},
\label{eq:workpower}
\end{eqnarray}
where $\langle \cdot \rangle$ denotes expectation value with respect to the state $\rho_S(t)$ of the system at time $t$.
The heat currents and the  {power} fulfill the first law of thermodynamics
\begin{equation}
\label{eq:firstlaw}
\dot{U}_S (t) = \dot{W}(t) + \sum_{\alpha=1}^{N_B} \dot{Q}_\alpha(t),    
\end{equation}
where we have defined the internal energy as
\begin{equation}
U_S (t)={\rm Tr} \left[H_S(t)\rho_S(t)\right].
\end{equation}
Notice that we follow the convention that work done on the system and heat flowing into the system are positive. 

Since we are dealing with linearly coupled harmonic oscillators, we find it convenient to describe the evolution of the system using the formalism of Gaussian continuous variable systems~ {\cite{BraunsteinRMP2005}}. 
To this end, we define a vector $R$ containing the system's position and momentum operators,
\begin{equation}
R = \{ x_1,p_1,x_2,p_2, \dots x_N, p_N\}^T.    
\end{equation}
In terms of $R$, we can express the system's Hamiltonian as a quadratic form,
\begin{equation}
H_S(t) = \frac{1}{2} R A_S(t) R^T,
\end{equation}
with
\begin{equation}
A_S(t) = B(t) + C(t),
\end{equation}
where
\begin{gather}
B(t) = \bigoplus_{i=1}^N \begin{pmatrix}
\omega_i^2(t) & 0\\
0 & 1
\end{pmatrix} ,\nonumber\\ \nonumber\\
C(t) = \begin{pmatrix}
0 & C_{12} & \ldots & C_{1N}\\
C_{21} & 0 & & \vdots\\
\vdots & & & \vdots \\
C_{N1} & C_{N2} & \ldots & 0
\end{pmatrix} ,\nonumber\\ \nonumber\\
C_{ij} = \begin{pmatrix}
\lambda_{i,j}\sqrt{\omega_{i}(t) \omega_{j}(t)} & 0\\
0 & \lambda_{i,j}\sqrt{\frac{1}{\omega_{i}(t) \omega_{j}(t)}}
\end{pmatrix}.
\end{gather}

The evolution of Gaussian states can be described completely by the first and second moments of $R$.
The first moments fulfill the linear differential equation
\begin{equation}
    \dot{\langle R\rangle} = D(t) \langle R\rangle,
\end{equation}
where we have defined $D(t)=S_N A_S(t)-K$, being
\begin{equation}
S_N = \bigoplus_{i=1}^N \left(\begin{array}{cc}0 & 1 \\-1 & 0\end{array}\right)
\end{equation}
the symplectic matrix and
\begin{equation}
K = \frac 12 \bigoplus_{i=1}^{N} \sum_{\alpha=1}^{N_B} g^2_{i,\alpha} \left(\begin{array}{cc}1 & 0 \\0 & 1\end{array}\right)
\end{equation}
the dissipation term.
The second moments can be described by the covariance matrix elements defined as
\begin{equation}
\label{eq:covdefinition}
\sigma_{kl} (t) = \frac{1}{2}\expval{R_k R_l + R_l R_k} - \expval{R_k}\expval{R_l}   .
\end{equation}
Using the master equation in Eq.~\eqref{eq:me}, it is possible to show that
the covariance matrix fulfills the equation (see, for example, Ref.~\cite{NicacioPRA2016}),
\begin{equation}
\label{eq:dotsigma}
\dot\sigma=D\sigma+\sigma D^T+T,
\end{equation}
where we have defined the noise matrix,
\begin{equation}
T = \frac 12\bigoplus_{i=1}^{N} \sum_{\alpha = 1}^{N_B} g^2_{i,\alpha}(2n_\alpha+1) \left(\begin{array}{cc}\frac{1}{\omega_{i}(t)} & 0 \\0 & \omega_{i}(t)\end{array}\right).
\end{equation}

\begin{figure*}[t]
\begin{centering}
\begin{tabular}{l l}
\includegraphics[width=0.75\columnwidth]{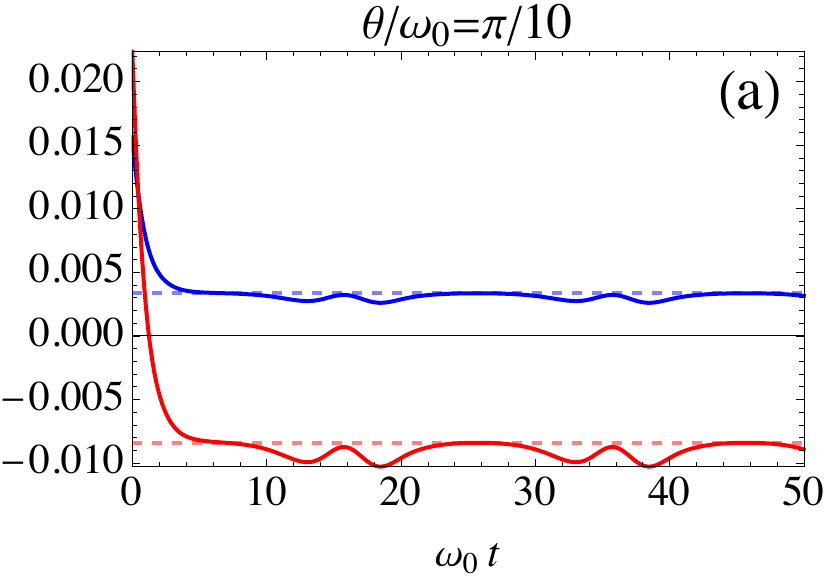}&
\includegraphics[width=1.05\columnwidth]{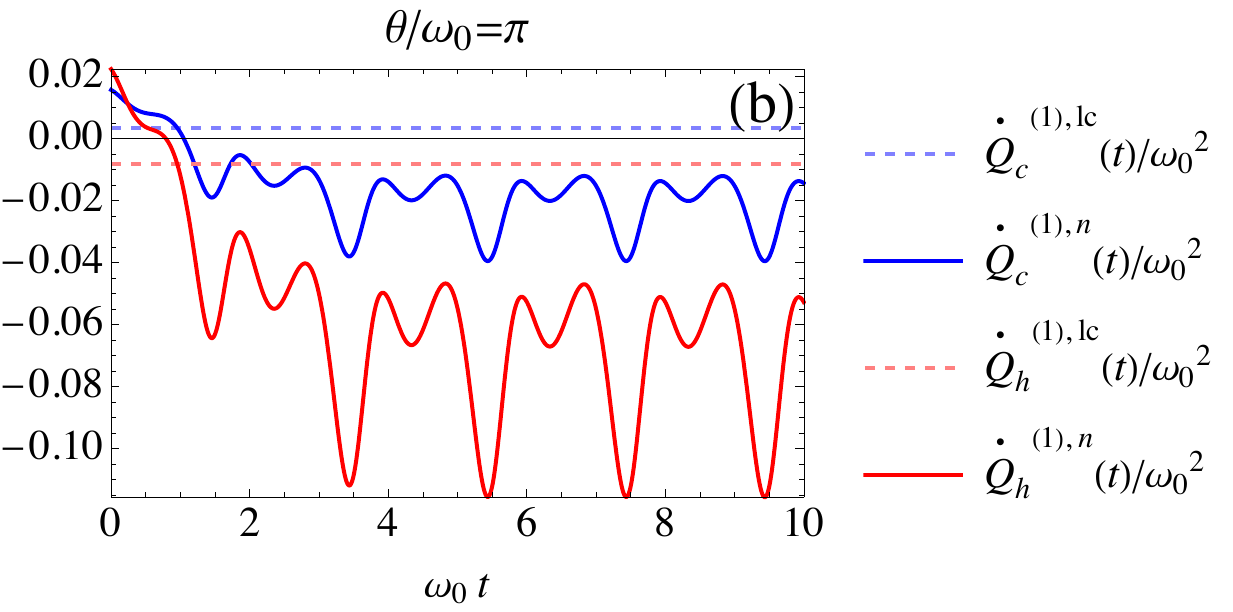}\\
\includegraphics[width=0.75\columnwidth]{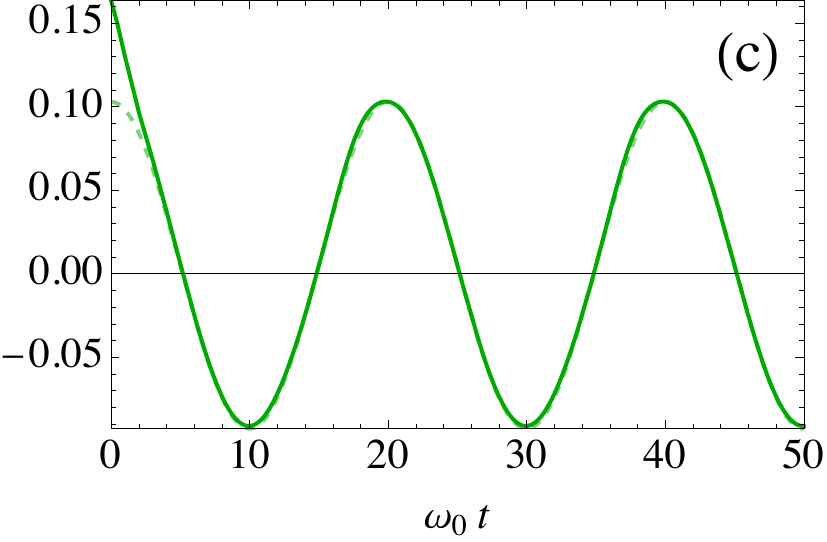}&
\includegraphics[width=1.05\columnwidth]{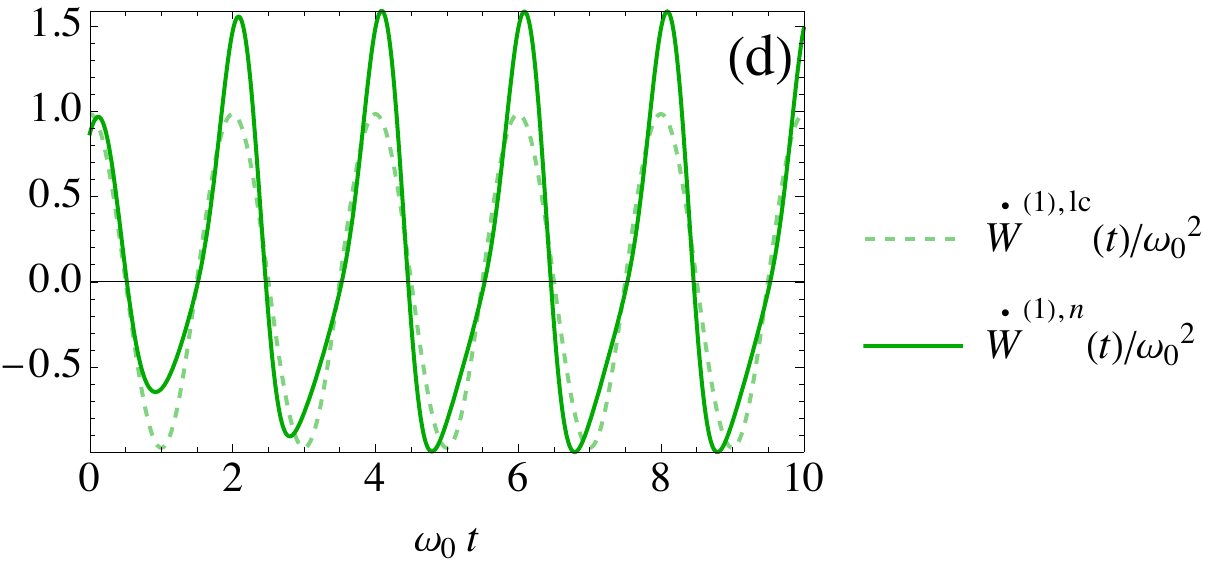}
\end{tabular}
\caption{The heat flows, (a) and (b), and  power, (c) and (d) for the one-oscillator system found by solving the master equation  {from Eq.~\eqref{eq:me}} numerically (superscript $n$) versus the ``limit cycle" solutions (superscript lc). We compare slow, (a) and (c), and fast driving, (b) and (d). The initial conditions used are given in Eq.~\eqref{eqinitialcons}. Parameters: 
$\delta \omega = \omega_0 /2,  \; g_c = g_h = \sqrt{\omega_0/2}, \; \beta_c = 10/\omega_0, \; \beta_h = 5/\omega_0, \; \Omega_c = \omega_0/5, \; \Omega_h = \omega_0/2$.}
\label{figlimitcycle}
\end{centering}
\end{figure*}

\section{One-Oscillator System}
\label{sec:cov1}

In this section, we showcase the results for the simplest example of one system oscillator, with driving frequency,
\begin{equation}
\omega(t) = \omega_0 + \delta\omega\sin(\theta t),
\end{equation}
in contact with two baths at different temperatures ($N=1, N_B=2$). For clarity of notation, we assume bath 1 to be the cold bath and denote its quantities with the subscript $c$, for example $n_1=n_c$ and $g_{1,1}=g_c$. Similarly, we assume bath 2 to be the hot bath and denote its quantities with the subscript $h$, for example, $n_2=n_h$ and $g_{1,2}=g_h$.

In this case, the master equation takes the simple form,
\begin{eqnarray}
\dot{\rho}_S(t)  &=& - \ii \comm{H_S(t)}{\rho_S(t)} \nonumber \\
& +& \gamma \bar n\left(a_1^\dagger \rho_Sa_1 -\frac{1}{2}\left \{a_1a_1^\dagger ,\rho_S\right\}\right)
\nonumber
\\
&+&\gamma(\bar n+1)\left(a_1 \rho_Sa_1^\dagger -\frac{1}{2}\left\{a_1^\dagger a_1 ,\rho_S\right\}\right),
\label{eq:me1}
\end{eqnarray}
where $\gamma=g^2_c+g^2_h$ is the total dissipation rate and
\begin{equation}
\label{eq:nbar}
\bar n = \frac{g^2_c n_c+g^2_h n_h}{g^2_c+g^2_h}
\end{equation}
 is the weighted average thermal occupation. Equation~\eqref{eq:me1} shows that the interaction of a single oscillator with two thermal baths is equivalent to that of one oscillator in contact with an effective bath with a weighted average temperature. An analogous result holds in the case of a larger number of baths. While this effective picture is true for the system dynamics, the individual heat currents $\dot Q_\alpha$ depend on the specific temperatures and couplings of the baths and are given by Eq.~\eqref{eq:Qdot}.

\begin{figure*}[t]
\begin{centering}
\includegraphics[width=0.6\columnwidth]{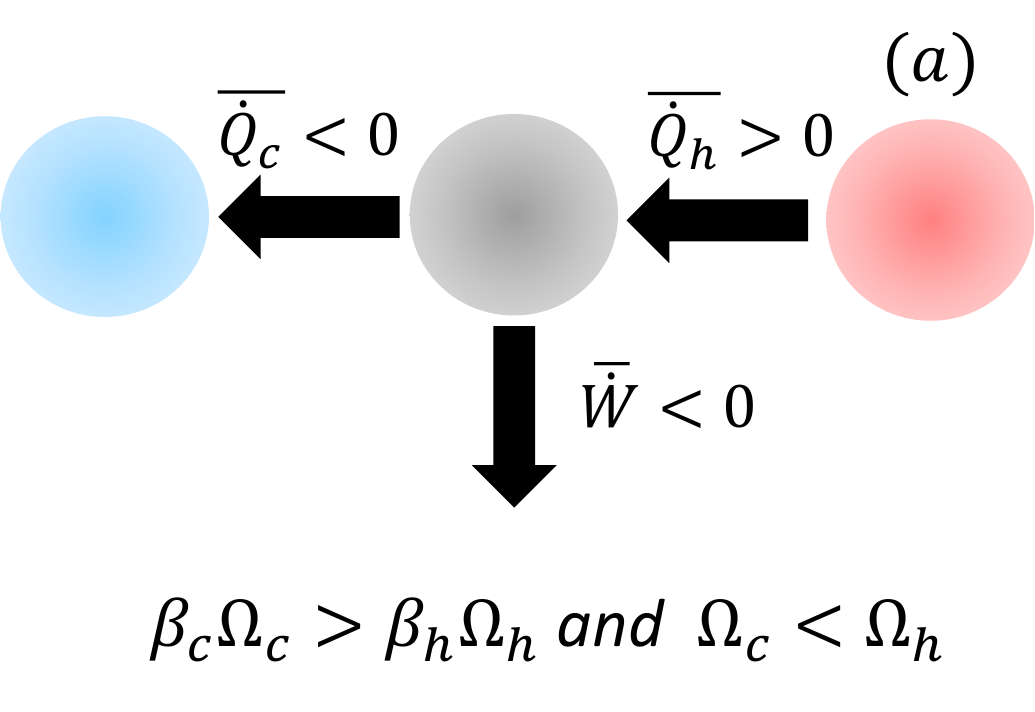}
\includegraphics[width=0.6\columnwidth]{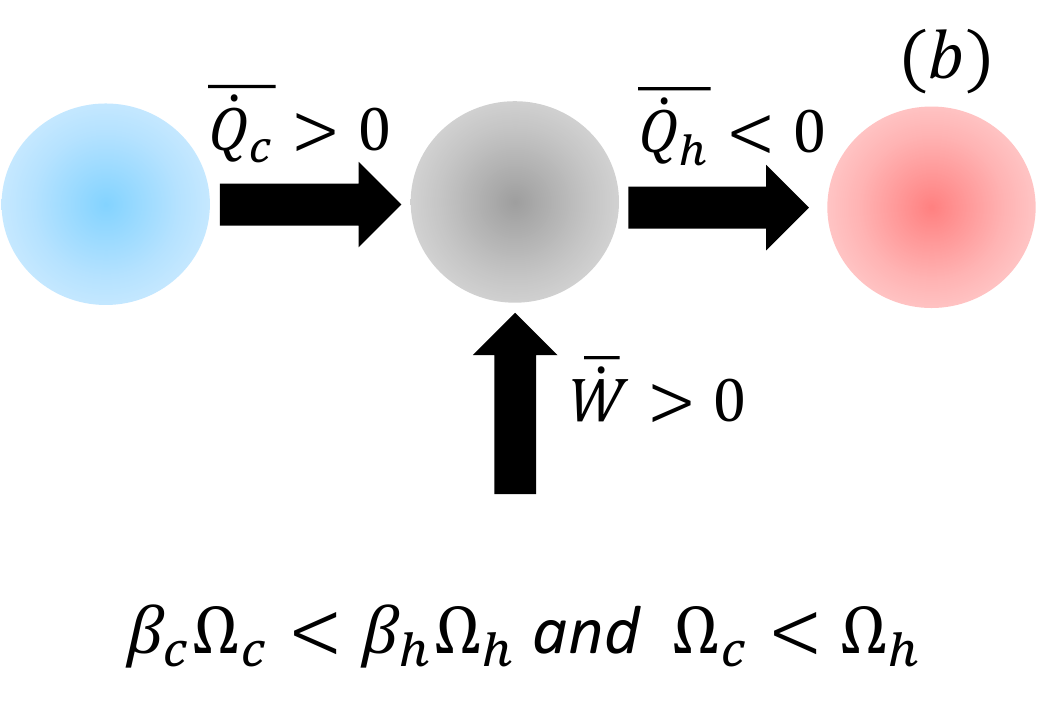}
\includegraphics[width=0.6\columnwidth]{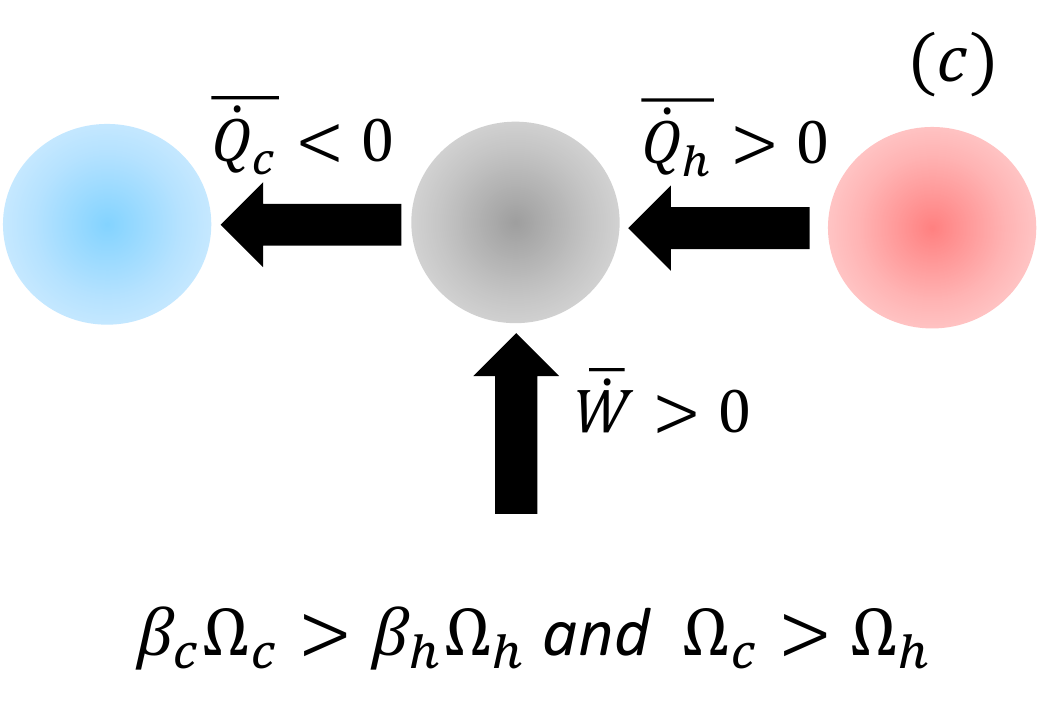}
\caption{Possible functionings of a single oscillator system  {where the bar represents the average over a period of oscillation}. (a) Engine, (b) refrigerator, and (c) accelerator. Below each scheme the conditions for the operating regime are shown.}
\label{fig:thermalmachines}
\end{centering}
\end{figure*}

Similarly, using the formalism introduced in Sec.~\ref{sec:model}, we can obtain the equivalent dynamical equation of a system made of one driven QHO in terms of the covariance matrix [see Eq.~\eqref{eq:dotsigma}],
\begin{equation}
{\sigma}(t) = 
\begin{pmatrix}
{\sigma}_{xx}(t) & {\sigma}_{xp}(t)\\
{\sigma}_{xp}(t) & {\sigma}_{pp}(t)
\end{pmatrix},
\end{equation}
where 
\begin{equation}
\begin{split}
\dot{\sigma}_{xx}(t) &= -\gamma\sigma_{xx}(t)  +\frac{\gamma(2\bar n+1)}{2 \omega(t)}  + 2\sigma_{xp}(t), \\
\dot{\sigma}_{xp}(t) &= - \gamma \sigma_{xp}(t) + \sigma_{pp}(t) - \sigma_{xx}(t) \omega^2(t),\\
\dot{\sigma}_{pp}(t) &=-\gamma \sigma_{pp}(t) + \frac{\gamma(2\bar n+1)\omega(t)}{2} -  2\sigma_{xp}(t) \omega^2(t).
\end{split}
\end{equation}
From Eq.~\eqref{eq:Qdot}, the heat flows can be calculated as
\begin{equation}\label{eqheatcovc}
\dot{Q}_\alpha^{(1)}(t) \! = \! -\frac{g_\alpha^2 \Omega_\alpha}{2 \omega(t)}\!\left[ \sigma_{pp}(t)\! + \! \omega^2(t)\sigma_{xx}(t) \! - \! \omega(t) (2n_\alpha+1) \right ],
\end{equation}
where $\alpha=c,h$
and the  {power} is
\begin{eqnarray}
\label{eq:work1}
\dot{W}^{(1)}(t)&=&-\frac{\gamma}{2}\sigma_{pp}(t)+\omega(t)\dot\omega (t)\sigma_{xx}(t)
\nonumber\\
&+&\frac{\gamma \omega(t)}{2}[2\bar n +1-\omega(t)\sigma_{xx}(t)] \nonumber
\\
&-& \dot{Q}_c^{(1)}(t) - \dot{Q}_h^{(1)}(t),
\end{eqnarray}
containing a term explicitly depending on the derivative  $\dot\omega (t)$ of the oscillator's frequency.

In the rest of this section, we are going to show a few examples of the thermodynamics of one oscillator.

\subsection{Slow driving}
\label{sec:slow2}
Let us first look at the slow driving case where the driving frequency is changing  slowly compared to the typical timescales of the system, $\theta_1\ll\{\gamma,\omega\}$. In this case, we can set $\dot{\sigma}_S(t) = 0$, assuming the system to instantaneously follow the steady state imposed by the instantaneous $\omega(t)$, and solve for $\sigma_{xx}(t), \sigma_{xp}(t)$, and $\sigma_{pp}(t)$. We find
\begin{eqnarray}
\sigma_{xx}(t) &=& \frac{2\bar n +1}{2\omega(t)}, \nonumber
\\
\sigma_{xp}(t) &=& 0, \nonumber 
\\
\sigma_{pp}(t) &=& \omega(t) \frac{2\bar n +1}{2}.
\end{eqnarray}
This state corresponds to a thermal state of a QHO with frequency $\omega(t)$ in equilibrium with an environment with thermal occupation $\bar n$.
Plugging these solutions into Eqs.~\eqref{eqheatcovc} and \eqref{eq:work1} for the heat flows and   {power}, we get
\begin{eqnarray}
\dot{Q}_c^{(1)}(t) &=& \frac{\Omega_c g_c^2 g_h^2 (n_c - n_h)}{g_c^2 + g_h^2} , \nonumber
\\
\dot{Q}_h^{(1)}(t) &=& \frac{\Omega_h g_c^2 g_h^2 (n_h - n_c)}{g_c^2 + g_h^2} , \nonumber \label{eqlceng}
\\
\dot{W}^{(1)}(t) &=& \frac{g_c^2 g_h^2 (\Omega_c - \Omega_h)(n_h - n_c)}{g_c^2 + g_h^2} 
+ \frac{2\bar n+1}{2}\dot\omega(t). \nonumber\\
\end{eqnarray}
Notice that under the assumption of slow driving, the heat currents are constant. This is due to the fact that instantaneously the system is in the steady state corresponding to the instantaneous value of the frequency $\omega(t)$.
Note that the second term in $\dot{W}(t)$ is proportional to $\dot{\omega}(t)$ and therefore, since $\omega(t)$ is periodic, it does not contribute to the average value. 

Figure \ref{figlimitcycle} compares the approximate ``limit cycle" solutions \eqref{eqlceng} with the exact numerical results from Eqs.~\eqref{eqheatcovc} and \eqref{eq:work1}. Choosing a modulation frequency $\theta$ small compared to $\omega$ and $\gamma$, as in the top and bottom left panels, we see that after a brief transient period that depends on the initial conditions, the heat flows and  {power} still oscillate slightly, but tend to the limit cycle solutions.

In Fig.~\ref{figlimitcycle}, we have assumed the system to be initially in a Gibbs state in thermal equilibrium at the cold inverse temperature $\beta_c$ and initial frequency $\omega(0)$, so that we have the initial conditions,
\begin{eqnarray}\label{eqinitialcons}
\sigma_{xx}(0) &=& \frac{2n_0 + 1}{2\omega(0)}, \nonumber\\
\sigma_{xp}(0) &=& 0,\nonumber\\
\sigma_{pp}(0) &=& \frac{\omega(0)(2n_0 + 1)}{2},
\end{eqnarray}
where
\begin{equation}
    n_0 = \frac{1}{e^{\beta_c \omega(0)} -1}.
\end{equation}

The slow driving solutions of Eq.~\eqref{eqlceng} allow us to interpret the functioning of the single oscillator system as a thermal device. Depending on the parameters $n_c,n_h,\Omega_c$, and $\Omega_h$, the device can function as an engine, a refrigerator or an accelerator. In the following we briefly illustrate these functionings.

\subsubsection{Engine}
To create an engine we need heat to flow from the hot bath, through the system and to the cold bath and in doing this, the system should have a certain power output. In order to achieve this functioning we require that  {$\overline{\dot{Q}^{(1)}_c} < 0, \; \overline{\dot{Q}^{(1)}_h} > 0$, and $\overline{\dot{W}^{(1)}} < 0$, where the bar indicates the average over a period of oscillation.} The corresponding operating mode is shown in Fig.~\ref{fig:thermalmachines}. From Eq.~\eqref{eqlceng},
this happens when $n_c < n_h$, i.e., when $\beta_h \Omega_h < \beta_c \Omega_c$, and $\Omega_c < \Omega_h$.

\subsubsection{Refrigerator}
For a refrigerator, we instead need a power input that enables heat to flow from the cold bath, through the system and to the hot bath. This functioning on average over a period of oscillation requires  {$\overline{\dot{Q}^{(1)}_c} > 0, \; \overline{\dot{Q}^{(1)}_h} < 0$, and $\overline{\dot{W}^{(1)}} > 0$} which, from Eq.~\eqref{eqlceng}, happens when $n_h < n_c$, i.e., when $\beta_h \Omega_h > \beta_c \Omega_c$, and $\Omega_c < \Omega_h$ as shown in Fig.~\ref{fig:thermalmachines}.

\subsubsection{Accelerator}\label{secheatpump}
An accelerator occurs when, like an engine, we have heat flowing from the hot bath to the system and from the system to the cold bath, i.e.,  {$\overline{\dot{Q}^{(1)}_c} < 0$ and $\overline{\dot{Q}^{(1)}_h}>0$}, the process being sped up, on average, by inputting power over a period of oscillation,  {$\overline{\dot{W}^{(1)}} > 0$}, as shown in Fig.~\ref{fig:thermalmachines}. To achieve this regime in Eq.~\eqref{eqlceng}, we require that $n_c < n_h$, meaning $\beta_c \Omega_c > \beta_h \Omega_h$, and $\Omega_h < \Omega_c$. \\

During a cycle, the signs of the thermodynamic quantities may change. For instance, while the average  {power} over a cycle may be negative, causing the system to operate as an engine overall, there could be instants during the cycle when the  {power} becomes positive so that the device temporarily switches to an accelerator.  

\begin{figure}[t]
\includegraphics[width=0.85\columnwidth]{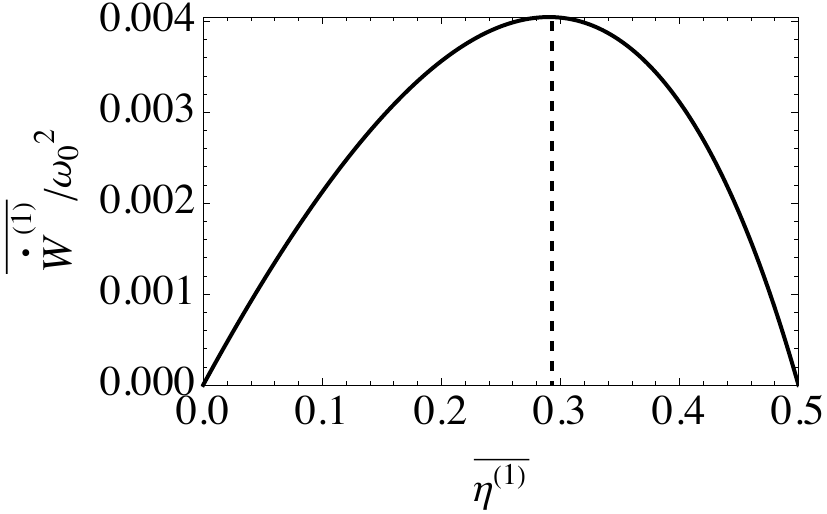}
\caption{Plot of the  {power} output against the average efficiency for the one-oscillator system, both quantities averaged over a period. The dashed line corresponds to the maximum  {power} which occurs at the Curzon-Ahlborn condition when $\Omega_h = \sqrt{\frac{\beta_c}{\beta_h}}\Omega_c$. Other parameters: $g_c = g_h = \sqrt{\omega_0/2}, \; \beta_c = 10/\omega_0, \; \beta_h = 5/\omega_0, \; \Omega_c = \omega_0/10$. 
}
\label{figmaxpow1}
\end{figure}
\begin{figure}[t]
\begin{centering}
\includegraphics[width=1\columnwidth]{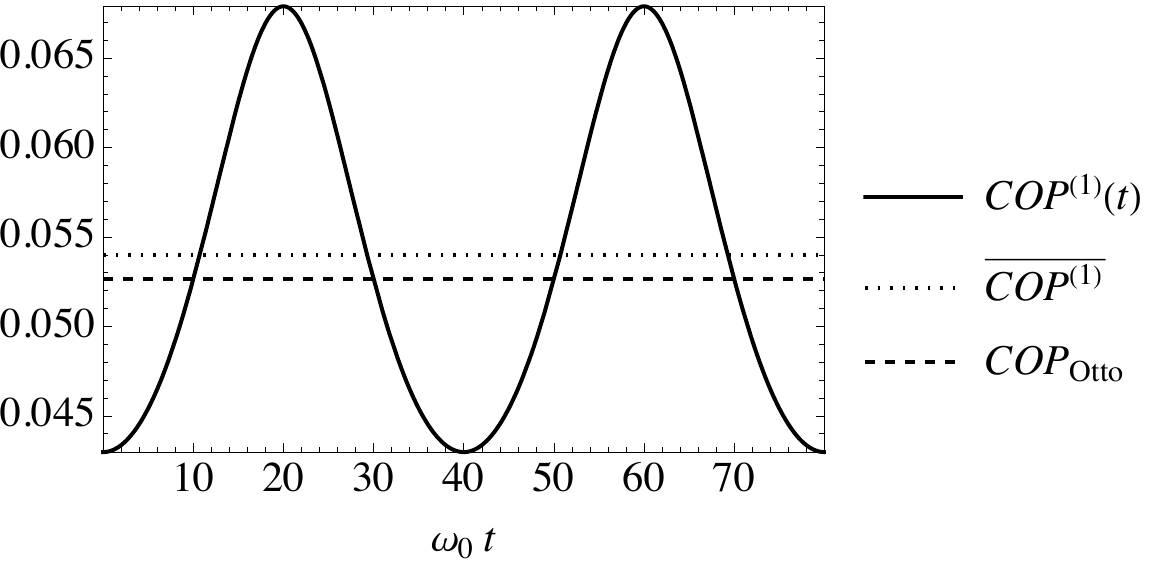}
\caption{ COP of a refrigerator for the one-oscillator system, its average over a period, and the Otto COP value plotted against time. Parameters: $\delta \omega = \omega_0/2 ,\; \theta/\omega_0 = \pi/20,\; g_c = g_h = \sqrt{\omega_0/2}, \; \beta_c = 10/\omega_0, \; \beta_h = 5/\omega_0, \; \Omega_c = \omega_0/10, \; \Omega_h = 2 \omega_0$.}
\label{fig:COP}
\end{centering}
\end{figure}

\subsubsection{Efficiency and coefficient of performance}\label{secefficiencycop}
The instantaneous efficiency of the engine can be defined as
\begin{equation}\label{eqefficiency}
\eta^{(1)} (t) = -\frac{\dot{W}^{(1)}(t)}{\dot{Q}^{(1)}_h(t)},
\end{equation}
with the condition $\dot{W}^{(1)}(t)<0$.
Plugging in the limit cycle solutions, we get
\begin{equation}\label{eq:eta1}
\eta^{(1)} (t) = \eta_{\rm Otto} + \delta \eta^{(1)}(t),
\end{equation}
where we have defined the efficiency of an Otto cycle operating between frequencies $\Omega_c$ and $\Omega_h$ as $\eta_{\rm Otto}=1-\Omega_c/\Omega_h$ and the time-dependent correction,
\begin{equation}
\delta\eta^{(1)}(t)=\frac{\dot{\omega}(t)(2\bar n+1)}{2\Omega_h(n_h-n_c)}\left(\frac{1}{g_c^2}+ \frac{1}{g_h^2}\right).
\end{equation}
Since $\omega(t)$ is a periodic function, the average of $\dot\omega$ over a period is zero and we obtain that the average efficiency over a period reduces to the Otto efficiency, i.e.,  {$\overline{\eta^{(1)}}=\eta_{\rm Otto}$} since  {$\overline{\delta\eta^{(1)}}=0$}.

For an engine, we are often interested in a  regime in which its power output is maximum and not when its efficiency is the highest. Since the average efficiency  {$\overline{\eta^{(1)}}=\eta_{\rm Otto}$} is equal to the 
 Otto value, we expect  the average efficiency at maximum power within these assumptions to be equal to the Curzon-Ahlborn value~\cite{Curzon1975},
\begin{equation}
\eta_{CA} = 1- \sqrt{\frac{\beta_h}{\beta_c}}.
\end{equation}
This is true for a system of one QHO and occurs when $\Omega_h = \sqrt{\frac{\beta_c}{\beta_h}}\Omega_c$. For one possible choice of parameters this is illustrated in Fig.~\ref{figmaxpow1}.

\begin{figure*}[t]
\begin{centering}
\includegraphics[height=3.5cm]{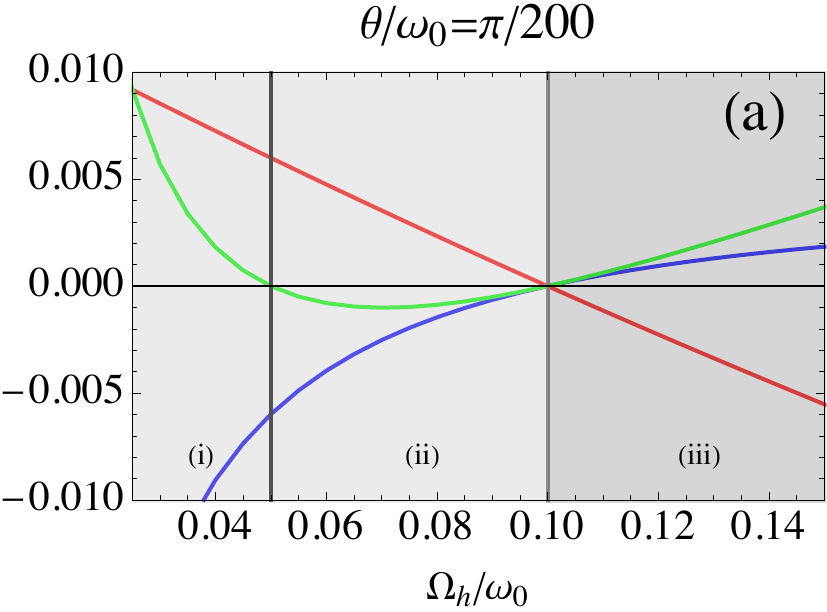}
\includegraphics[height=3.5cm]{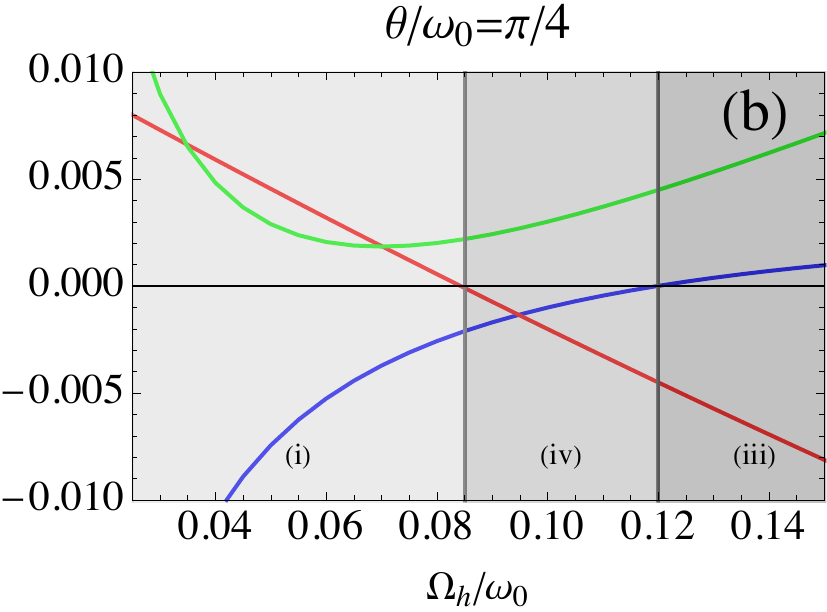}
\includegraphics[height=3.6cm]{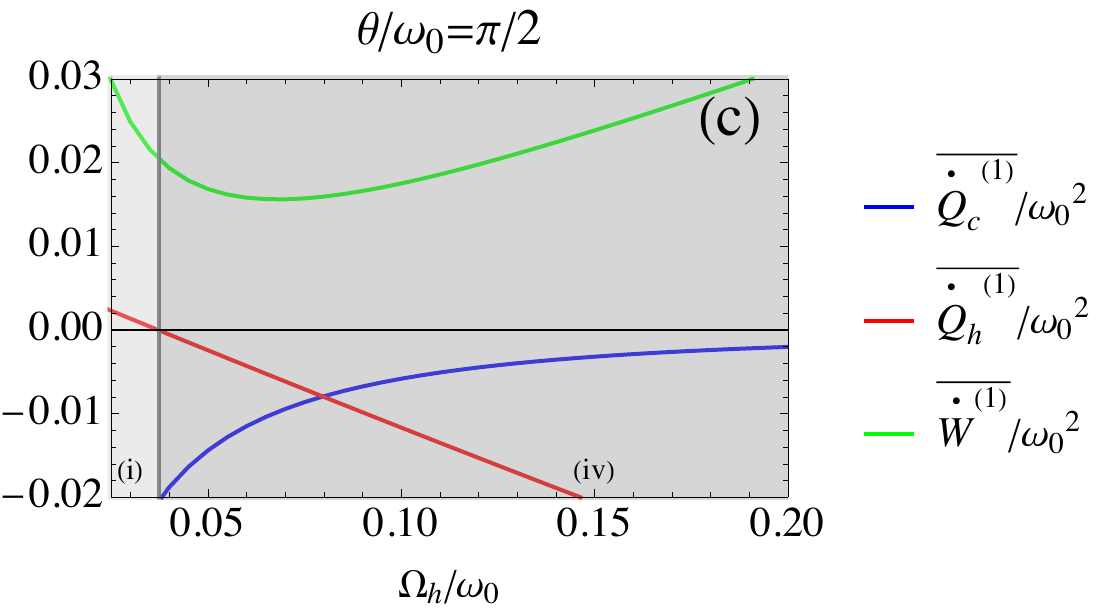}
\caption{ Plots of average heat flows and  {power} for the one-oscillator system as a function of $\Omega_h$.
(i) Accelerator, (ii) engine, (iii) refrigerator, and (iv) dissipator.
(a) Slow driving: $\theta/\omega_0 = \pi/200$.
(b) Increased driving: $\theta/\omega_0 = \pi/4$.
(c) Fast driving: $\theta/\omega_0 = \pi/2$.
The initial conditions are given in Eq. \eqref{eqinitialcons}. Parameters: $\delta \omega = \omega_0/2 ,\;  g_c = g_h = \sqrt{\omega_0}/2, \; \beta_c = 20/\omega_0, \; \beta_h = 10/\omega_0, \; \Omega_c = \omega_0/20$.
}
\label{fig:NAfunctions1}
\end{centering}
\end{figure*}

In the refrigerator regime, the functioning of the thermal device can be assessed by the coefficient of performance (COP) of the refrigerator defined as
\begin{equation}\label{eqcopr}
COP^{(1)}(t) = \frac{\dot{Q}^{(1)}_c(t)}{\dot{W}^{(1)}(t)},
\end{equation}
with $\dot{Q}^{(1)}_c(t)>0$.
Plugging in the limit cycle solutions, we get
\begin{equation}\label{eqeff1}
\frac{1}{COP^{(1)}(t)} = \frac{1}{COP_{\rm Otto}} + \frac{1}{\delta COP^{(1)}(t)},
\end{equation}
where we have defined $COP_{\rm Otto}=\Omega_c/(\Omega_h-\Omega_c)$ and
\begin{equation}
 \frac{1}{\delta COP^{(1)}(t)}= -\frac{\Omega_h}{\Omega_c} \delta \eta^{(1)}(t).
\end{equation}
Since $\delta COP^{(1)}(t)$ oscillates, between positive and negative values, the instantaneous COP can be larger than the Otto value $COP_{\rm Otto}$. Even more, due to the non-linear relation in which $\delta COP^{(1)}(t)$ enters the expression of the COP, the average  {$\overline{COP^{(1)}}$} can be larger than $COP_{\rm Otto}$ as shown in Fig.~\ref{fig:COP}. 
On the other hand, the ratio of the average of the heat flow to the cold bath and the average of  {power} is equal to the Otto value $COP_{\rm Otto}$. 
Note that in Fig.~\ref{fig:COP} the parameters have been chosen so that the  {power} remains positive throughout each cycle allowing the system to operate as a refrigerator at all times.
These results show that, although on average (when taking the ratio of the averages of heat flow and power independently) the refrigerator with modulated driving performs as efficiently as an Otto refrigerator, the instantaneous coefficient of performance can in fact surpass it.

\subsection{Fast driving}

We now analyze what happens in the fast regime when the driving modulation frequency $\theta$ is comparable to the typical timescale of the system as dictated by $\gamma$ or $\omega_0$. In this regime, we can no longer assume $\dot{\sigma}_S(t) = 0$, so we must solve the master equation directly by numerical integration. We then use the resulting values of the covariance matrix to obtain the relevant thermodynamic quantities. 

After an initial transient period, we find that the heat currents always tend to some limit cycle in which the functions become periodic even though not with a simple sinusoidal dependence, as seen in the top and bottom-right panel of Fig.~\ref{figlimitcycle}. This is the result of the interplay of the oscillator's free evolution, the modulation of its frequency, and the dissipation induced by the baths. We integrate over a period to find the average values of heat flows and  {power}, which are plotted in Fig.~\ref{fig:NAfunctions1} as a function of $\Omega_h$ for increasing values of $\theta$. 

\begin{figure}[b]
\begin{centering}
\includegraphics[width=0.9\columnwidth]{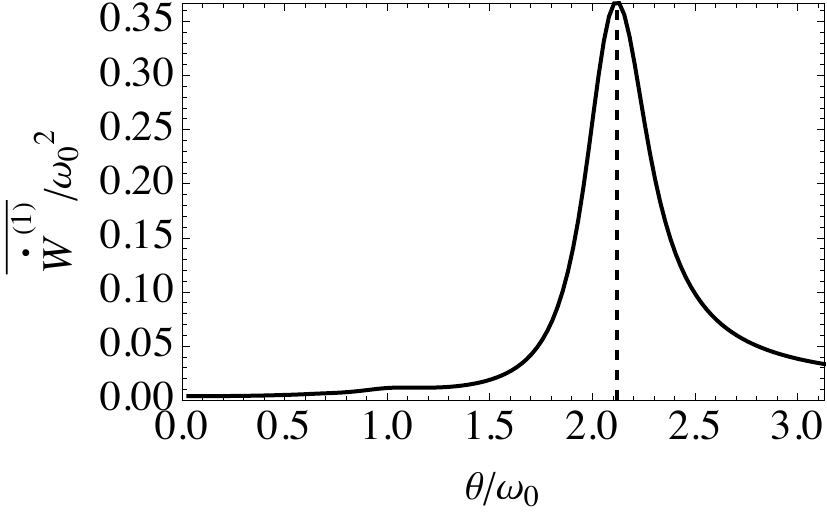}
\caption{Average power over a period for the one-oscillator system as a function of the modulation frequency $\theta$. There is a particular value of $\theta$ ($\theta/\omega_0 \approx 2.12 $, indicated by a vertical dashed line) for which the average  {power} injected over a cycle reaches a maximum. Parameters: $ \delta \omega = \omega_0/2 ,\; g_c = g_h = \sqrt{\omega_0}/2, \; \beta_c = 20/\omega_0, \; \beta_h = 10/\omega_0, \; \Omega_c = \omega_0/20, \;\Omega_h = 3\omega_0 /20$. }
\label{figresonanttheta}
\end{centering}
\end{figure}

For slow driving ($\theta/\omega_0 = \pi/200$) there is a value of $\Omega_h$ at which the heat flows and  {power} all go to zero. This is the Carnot point and occurs when $\frac{\beta_c}{\beta_h} = \frac{\Omega_h}{\Omega_c} $.
We find that the average dissipated  {power} increases with faster driving while the cooling power decreases, meaning that the COP for the refrigerator  decreases. Also, with faster driving the  {power} remains positive for all values of $\Omega_c$ and $\Omega_h$ so it is no longer possible for the system to operate as an engine. In particular, for larger $\theta$ we do find that the  {power} remains positive while both heat flows become negative. This regime corresponds to a dissipator:  {the increasingly fast driving results in a large injection of power into the system and is transformed into heat which is dissipated in both the cold and hot reservoirs. A dissipator may be used to dissipate heat generated in a mechanical or electronic device. Alternatively, a dissipator can be used to warm another system, hence the name heater.} The maxima of the dissipated heats $|\dot Q^{(1)}_{c}|$ and $|\dot Q^{(1)}_{h}|$ correspond to the maximum of the injected  {power} $\dot W^{(1)}$ and arise for a ``resonant" value of $\theta$ for which the conversion work to heat induced by the driving is more effective. This is seen in Fig. \ref{figresonanttheta}. Notice that as the process is dissipative, there is no thermodynamic constraint limiting the amount of heat produced. 

\section{Two Oscillators}
\label{sec:twoosc}
We now extend our study to a system of two driven QHO's ($N=2$), with driving frequencies $\omega_1(t)$ and $\omega_2(t)$, respectively, described by the Hamiltonian of Eq.~\eqref{eq:HS} with  {$\lambda_{12}=\lambda$}. We assume each QHO to be coupled to only one reservoir ($N_B=2$): oscillator 1 (2) is coupled to a cold (hot) reservoir characterized by frequency $\Omega_c \; (\Omega_h)$ and thermal occupation $n_c \; (n_h)$.  {It follows that in Eq.~\eqref{eq:HI}, $g_{1, 2}= g_{2, 1}=0$, while we set $g_c=g_{1,1}$ and $g_h=g_{2, 2}$.}
Our system can now be described by a 4 by 4 covariance matrix defined in Eq.~\eqref{eq:covdefinition} with matrix entries
\begin{equation}
{\sigma}_S(t) = 
\begin{pmatrix}
{\sigma}_{x_1x_1}(t) & {\sigma}_{x_1p_1}(t) & {\sigma}_{x_1x_2}(t) & {\sigma}_{x_1p_2}(t)\\
{\sigma}_{x_1p_1}(t) & {\sigma}_{p_1p_1}(t) & {\sigma}_{x_2p_1}(t) & {\sigma}_{p_1p_2}(t)\\
{\sigma}_{x_1x_2}(t) & {\sigma}_{x_2p_1}(t) & {\sigma}_{x_2x_2}(t) & {\sigma}_{x_2p_2}(t)\\
{\sigma}_{x_1p_2}(t) & {\sigma}_{p_1p_2}(t) & {\sigma}_{x_2p_2}(t) & {\sigma}_{p_2p_2}(t)
\end{pmatrix}.
\end{equation}
Just as in Sec.~\ref{sec:cov1}, we can find the heat currents 
\begin{eqnarray}
\dot{Q}^{(2)}_c &=& -\frac{g_c^2 \Omega_c}{2 \omega_1} \left[ \sigma_{p_1p_1} - \omega_1 (1 + 2 n_c) + \sigma_{x_1x_1} \omega^2_1 \right],
\\
\dot{Q}^{(2)}_h &=& -\frac{g_h^2 \Omega_h}{2 \omega_2} \left[( \sigma_{p_2p_2} - \omega_2 ( 1 + 2 n_h) + \sigma_{x_2x_2} \omega^2_2 \right],
\end{eqnarray}
which are extensions of Eq.~\eqref{eqheatcovc}. We have omitted the explicit time-dependence.
The expression for the  {power}, corresponding to Eq.~\eqref{eq:workpower}, in terms of the covariance matrix entries, is quite lengthy and reported in Appendix ~\ref{sec:appa}. As a check of the calculations, we have verified the first law of thermodynamics expressed in Eq.~\eqref{eq:firstlaw}.

\begin{figure}[t]
\begin{centering}
\includegraphics[width=1\columnwidth]{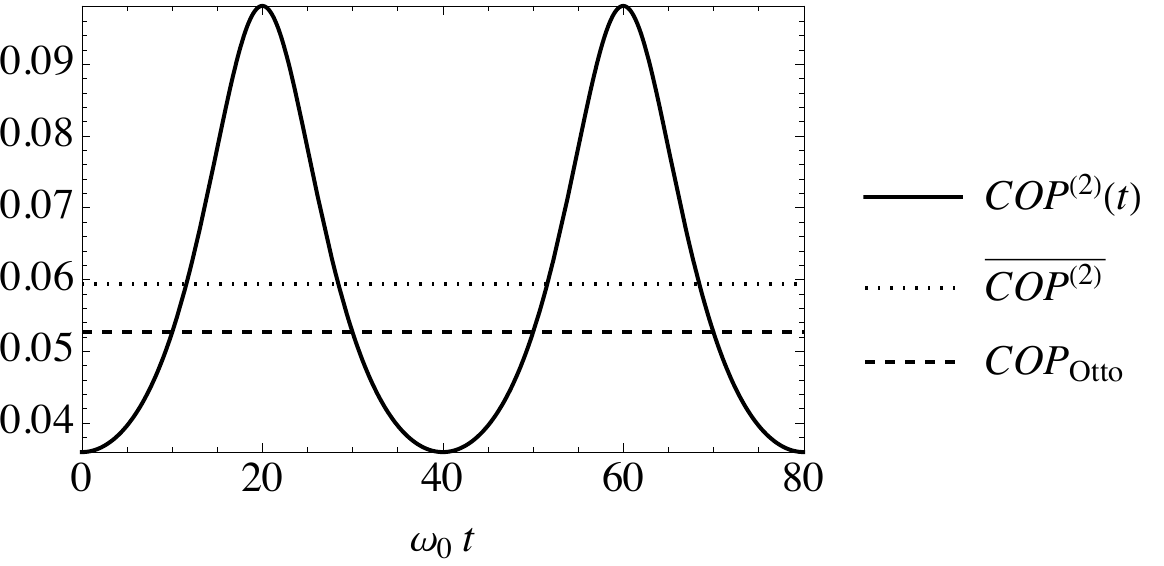}
\caption{ COP of a refrigerator made of two QHO's in the slow driving regime plotted as a function of time (solid line) compared with its average over a period (short-dashed) and the Otto value (long-dashed). Parameters: $\delta \omega = \omega_0/2 ,\; \theta/\omega_0 = \pi/20,\; g_c = g_h = \sqrt{\omega_0/2},\; \lambda =  \sqrt{2} \omega_0, \; \beta_c = 10/\omega_0, \; \beta_h = 5/\omega_0, \; \Omega_c = \omega_0/10, \; \Omega_h = 2 \omega_0$.}
\label{fig:COP2}
\end{centering}
\end{figure}

\subsection{Slow driving}
In the slow driving case, solving $\dot\sigma_S(t) = 0$ we can find analytically the  solution for the time-dependence of the covariance matrix in the limit cycle. The explicit form is not reported as it is quite lengthy. Using this expression, we obtain the heat currents,
\begin{equation}
\begin{split}
\dot{Q}^{(2)}_c &= \frac{4 \Omega_c \lambda^2 g_c^2 g_h^2 \left(g_c^2 + g_h^2\right) \left(n_c - n_h\right) }{\left(g_c^2 + g_h^2\right)^2 \left(4 \lambda^2 + g_c^2 g_h^2\right) + 4 g_c^2 g_h^2 [\omega_1 - \omega_2]^2},
\\
\dot{Q}^{(2)}_h &= - \frac{4 \Omega_h \lambda^2 g_c^2 g_h^2 \left(g_c^2 + g_h^2\right) \left(n_c - n_h\right) }{\left(g_c^2 + g_h^2\right)^2 \left(4 \lambda^2 + g_c^2 g_h^2\right) + 4 g_c^2 g_h^2 [\omega_1 - \omega_2]^2}.
\end{split}
\end{equation}
Again, the expression for the  {power} is quite lengthy and reported in Appendix \ref{sec:appa}. 

Notice that if $\omega_1(t) \neq \omega_2(t)$, due to the term proportional to $[\omega_1(t) - \omega_2(t)]^2$ in the denominator, the values of $\dot{Q}^{(2)}_c(t)$ and $\dot{Q}^{(2)}_h(t)$ [and indeed also of $\dot{W}^{(2)}(t)$] always have a smaller magnitude than when we have $\omega_1(t) = \omega_2(t)$. Therefore, from now on, we will assume that both QHO's have the same frequency $\omega_1(t)=\omega_2(t)=\omega(t)$ (and hence $\theta_1=\theta_2=\theta$). Under these assumptions, the expressions of heat currents simplify to
\begin{eqnarray}\label{eq:heatflow2a}
\dot{Q}^{(2)}_c = \frac{ \Omega_c  g_c^2 g_h^2  \left(n_c - n_h\right) }{\left(g_c^2 + g_h^2\right) \left(1 + \frac{g_c^2 g_h^2}{4 \lambda^2}\right)}=\frac{\dot{Q}_c^{(1)}}{1 + \frac{g_c^2 g_h^2}{4 \lambda^2}},    
\\
\dot{Q}_h^{(2)} = - \frac{ \Omega_h  g_c^2 g_h^2  \left(n_c - n_h\right) }{\left(g_c^2 + g_h^2\right) \left(1 + \frac{g_c^2 g_h^2}{4\lambda^2}\right)}=\frac{\dot{Q}_h^{(1)}}{1 + \frac{g_c^2 g_h^2}{4 \lambda^2}},  \label{eq:heatflow2b}
\end{eqnarray}
where in the last equations we have used the expression for the single oscillator heat currents of Eq.~\eqref{eqlceng}.
 {When the working substance is made of more than two linearly coupled resonant oscillators, with the same inter-oscillator coupling, the heat flows and the non-oscillating part of the power  coincide with the case with only two oscillators as in Eqs.~\eqref{eq:heatflow2a} and \eqref{eq:heatflow2b}, see Ref.~\cite{AsadianPRE2013}}.

\begin{figure}[t]
\begin{centering}
\includegraphics[width=1\columnwidth]{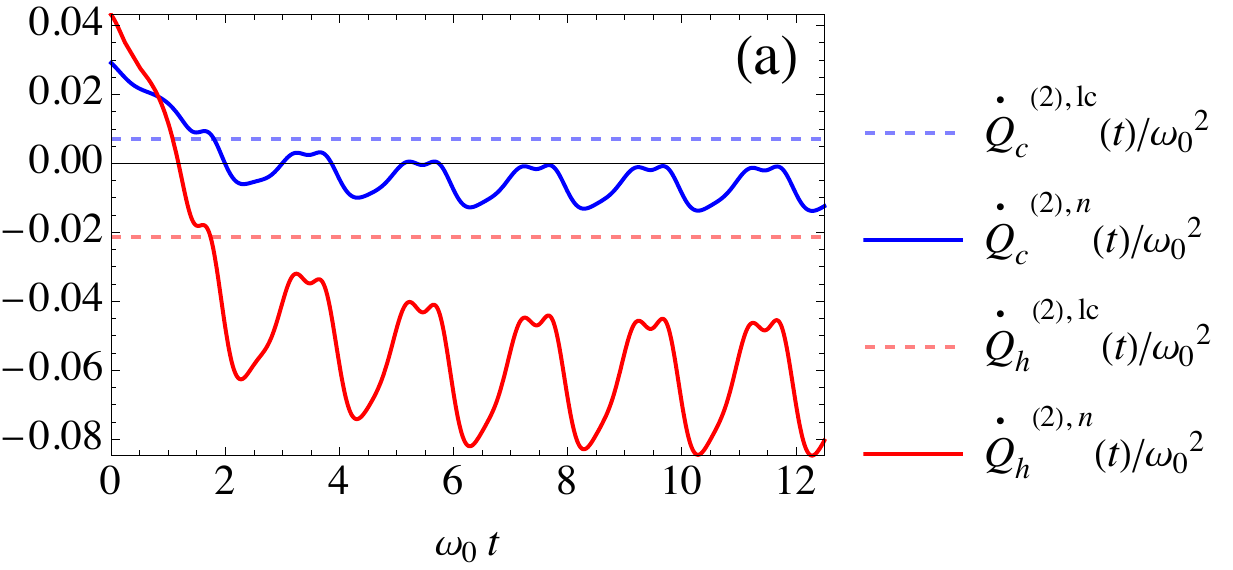}\\
\includegraphics[width=1\columnwidth]{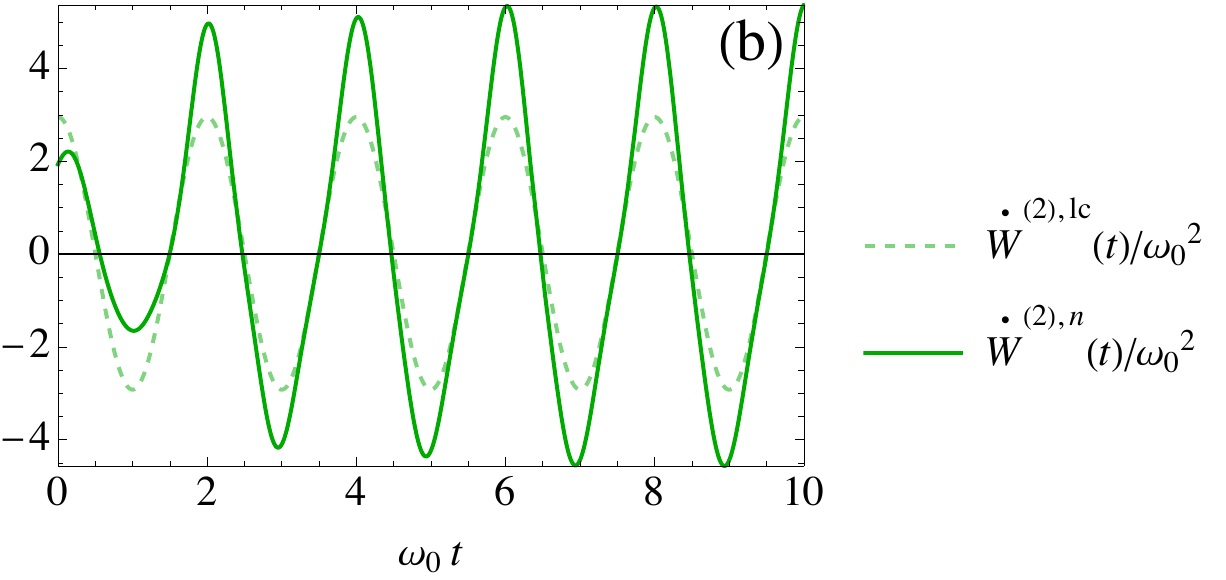}
\caption{Plots of the heat flows (a) and  {power} (b) for the two-oscillator system with fast driving (solid lines) compared to their values in the slow driving regime (dashed lines). Both oscillators have the same initial conditions as in Eq.~\eqref{eqinitialcons}. Parameters: $ \delta \omega = \omega_0 / 2, \; \theta/\omega_0 = \pi, \;g_c = g_h = \sqrt{\omega_0/2},\; \lambda =  \sqrt{2} \omega_0, \; \beta_c = 10/\omega_0, \; \beta_h = 5/\omega_0, \; \Omega_c = \omega_0/10, \; \Omega_h = 3\omega_0/10$. }
\label{fig:NA2}
\end{centering}
\end{figure}

\begin{figure*}[t]
\begin{centering}
\includegraphics[height=3.5cm]{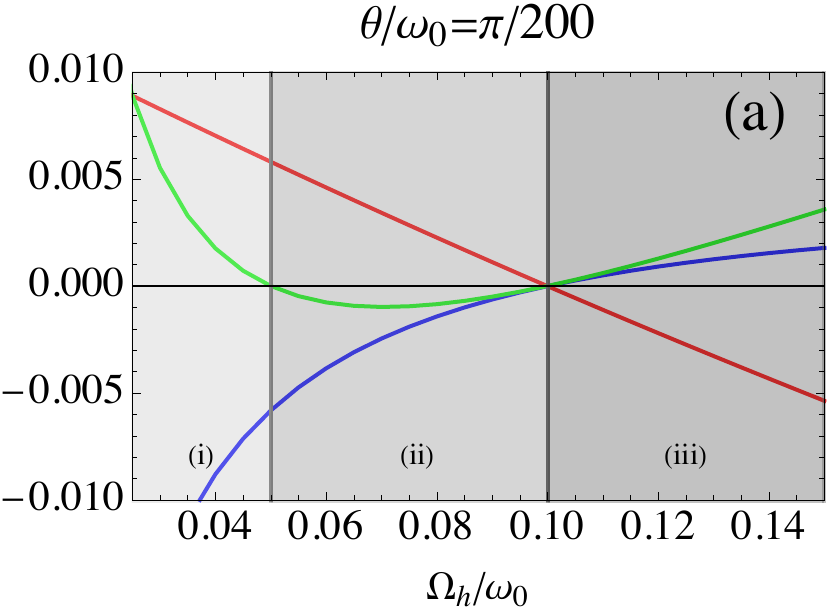}
\includegraphics[height=3.5cm]{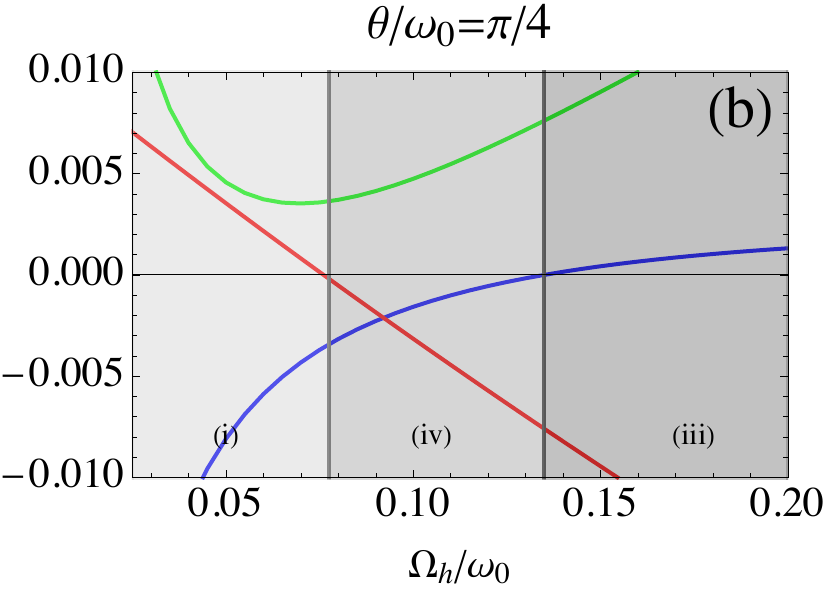}
\includegraphics[height=3.6cm]{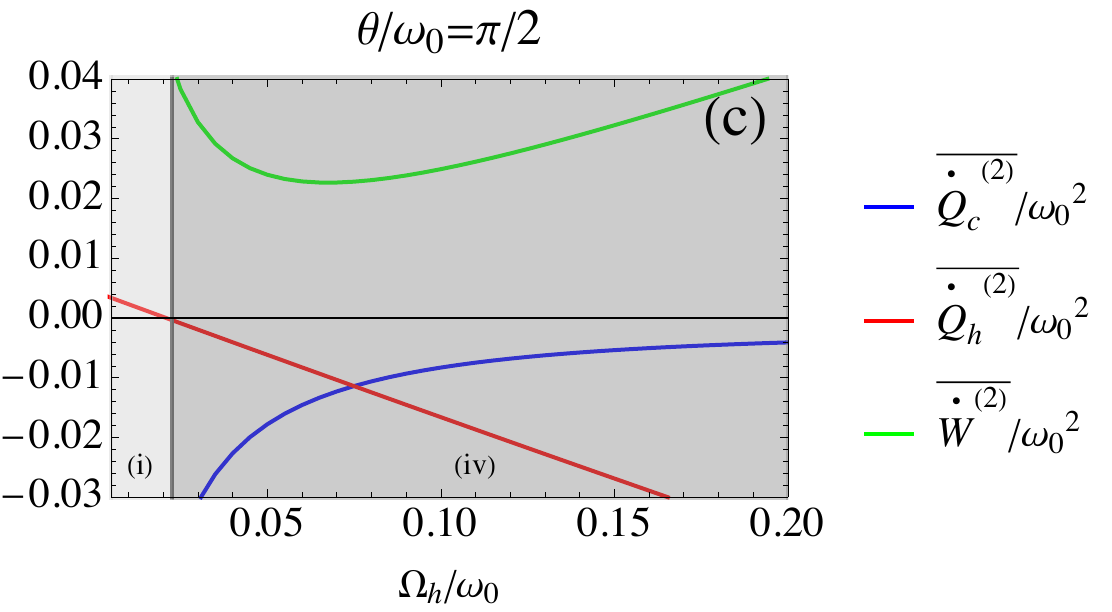}
\caption{Plots of average heat flows and  {power} for the two-oscillator system as a function of $\Omega_h$.
(i) Accelerator, (ii) engine, (iii) refrigerator, and (iv) dissipator.
(a) Slow driving: $\theta/\omega_0 = \pi/200$.
(b) Increased driving: $\theta/\omega_0 = \pi/4$.
(c) Fast driving: $\theta/\omega_0 = \pi/2$.
The initial conditions are given in Eq. \eqref{eqinitialcons}. Parameters: $\delta \omega = \omega_0/2 ,\;g_c = g_h = \sqrt{\omega_0}/2,\; \lambda =  \omega_0/\sqrt{ 2},\; \beta_c = 20/\omega_0, \; \beta_h = 10/\omega_0, \; \Omega_c = \omega_0/20$.
}
\label{fig:NAfunctions2}
\end{centering}
\end{figure*}
As the coupling strength, $\lambda$, between the two system oscillators increases, the heat flows and power tend to the same expressions for the one-oscillator system,
\begin{eqnarray}
\lim_{\lambda \rightarrow \infty} \dot{Q}^{(2)}_c(t) &=& \frac{\Omega_c g_c^2 g_h^2  (n_c - n_h) }{g_c^2 + g_h^2} =  \dot{Q}_c^{(1)}(t),
\\  
\lim_{\lambda \rightarrow \infty} \dot{Q}^{(2)}_h(t) &=& \frac{\Omega_h g_c^2 g_h^2  (n_h - n_c) }{g_c^2 + g_h^2} =  \dot{Q}_h^{(1)}(t).
\end{eqnarray}
This result can be understood because in this regime, the two oscillators are so strongly coupled that only the center of mass effectively interacts with the two baths and is affected by the driving. The relative motion mode, on the other hand, stiffens, being characterized by a very large effective frequency which  increases with  $\lambda$.
In this limit, the expression for the power also reduces to the corresponding expression for one oscillator,
\begin{eqnarray}
\lim_{\lambda \rightarrow \infty} \dot{W}^{(2)}(t) &=& 
\frac{g_c^2 g_h^2 (\Omega_c - \Omega_h)(n_h - n_c)}{ (g_c^2 + g_h^2)} \nonumber\\
&+& \frac{\left[ g_c^2(2n_c +1) + g_h^2(2n_h +1)\right] \dot{\omega}(t)}{2 (g_c^2 + g_h^2)} \nonumber\\
&= & \dot{W}^{(1)}(t).
\end{eqnarray}

Under the assumption $\omega_1(t)=\omega_2(t)$, using the approach reported in Appendix \ref{sec:appa}, it is easy to see that the conditions required for the system to operate on average as an engine, a refrigerator or an accelerator are the same as those in Fig.~\ref{fig:thermalmachines} for any value of $\lambda$.

One may use Eq.~\eqref{eqefficiency} to show that the efficiency for this system of two quantum harmonic oscillators is
\begin{equation}
\eta^{(2)}(t) = \eta_\text{Otto} + \delta\eta^{(2)}(t),  
\end{equation}
where
\begin{equation}
\delta\eta^{(2)}(t) =
\frac{\gamma (1+n_c+n_h)}{4\lambda^2 \Omega_h (n_c-n_h)}\dot{\omega}(t) + 2 \delta \eta^{(1)}(t).
\end{equation}
Since  $\delta\eta^{(1)}(t)$, and hence $\delta\eta^{(2)}(t)$, is proportional to $\dot{\omega}(t)$, the average efficiency over a period reduces to  {$\overline{\eta^{(2)}} = \eta_\text{Otto}$}, as in the case of one QHO.  

The COP of a refrigerator may be found from Eq.~\eqref{eqcopr} satisfying
\begin{equation}
\frac{1}{COP^{(2)}(t)} = \frac{1}{COP_{\text{Otto}}} + \frac{1}{\delta COP^{(2)}(t)},
\end{equation}
where
\begin{equation}
\frac{1}{\delta COP^{(2)}(t)} = -\frac{\Omega_h}{\Omega_c} \delta \eta^{(2)}(t).
\end{equation}
In Fig.~\ref{fig:COP2} the parameters defining the conditions of the baths, driving of the system, and couplings of the system to the bath are the same as in Fig.~\ref{fig:COP}, with the addition of the coupling between the two system oscillators, $\lambda$. As in the case of Fig.~\ref{fig:COP}, here the system operates as a refrigerator at all times. Comparing these two plots we see that the instantaneous COP reaches a higher peak in Fig.~\ref{fig:COP2} and consequently a higher average over a cycle.

\subsection{Fast driving}
Let us now pass to the case of fast driving where the condition $\dot\sigma_S =0$ does not hold. We continue to assume the resonant condition $\omega_1(t) = \omega_2(t) = \omega(t)$. Solving the master equation numerically we find that, as with the one QHO system, the heat flows and  {power} reach a limit cycle after a brief transient period, as shown in Fig.~\ref{fig:NA2}.
Plots of the average heat flows and  {power} over a period and their corresponding regimes are shown in Fig.~\ref{fig:NAfunctions2}. They look very similar to the corresponding graphs for the system of one QHO in Fig.~\ref{fig:NAfunctions1}. However, note that for a two-QHO system with fast driving ($\theta/\omega_0 = \pi/2$) the  {power} is larger while the system is functioning as a dissipator.

\section{Squeezed Bath}
\label{sec:squeezed}

In this section, we consider the extension of our model in which one of the environments is composed  of squeezed thermal ancillas. These can be prepared by applying the single-mode squeezing operator defined as
\begin{equation}
S(\xi) = \exp(\frac{1}{2}\xi^* b_{\alpha,j}^2 - \frac{1}{2}\xi {b_{\alpha,j}^{\dagger}}^2),  
\end{equation}
where $\xi = r e^{i\phi} $ is a complex parameter which depends on two real values: the squeezing magnitude $r\ge 0$ and the squeezing phase $\phi$. 
For simplicity, we will look at the case where $\phi = 0$, corresponding to the squeezing of the position-like quadrature and anti-squeezing of the momentum-like quadrature. 
As we will see shortly, squeezing effectively raises the temperature of the bath, so for convenience we will only apply squeezing to the hotter bath labeled 2. This leads to the following average occupation in the thermal squeezed state:
\begin{eqnarray}
n_h^{(\text{eff})} &=&{\rm Tr} \left[b^\dagger_{2,j} b_{2,j} S(\xi) \rho_{B,2}S^\dagger(\xi)\right] \nonumber
\\
&=& \frac {(1+2n_h)\cosh 2r - 1}{2}. \label{eq:nheff}
\end{eqnarray}
Using the fact that $\cosh 2r \ge 1$ we find $n_h^{(\text{eff})} \ge n_h$ with the equality obtained only for $r=0$.  

For a system of two oscillators, we follow the same method as in Sec.~\ref{sec:model}---setting $\omega_1(t) = \omega_2(t) = \omega(t)$---to get the master equation and energy flows.  All the expressions for the heat currents and power are formally identical to those used in Sec.~\ref{sec:twoosc} with the replacement $n_h\to n_h^{(\text{eff})}$.

As a consequence, at the thermodynamic level, the only modification that the squeezing introduces is a higher temperature, and thus energy, in the corresponding bath. This is connected to the known fact that a thermal state minimizes the energy for fixed entropy. The squeezed thermal state is characterized by the same entropy but higher energy corresponding to an effective larger temperature. 

To find the effective temperature we can use
\begin{equation}
n_h^{(\text{eff})} = \Big(e^{\beta_h^{(\text{eff})}\Omega_h} - 1  \Big)^{-1},
\end{equation}
where $n_h^{(\text{eff})} $ is defined in Eq.~\eqref{eq:nheff}.
Solving for $\beta_h^{(\text{eff})}$ we get
\begin{equation}
\beta_h^{(\text{eff})} = \frac{1}{\Omega_h} \log\Bigg[\frac{\tanh^2(r) +\exp(\beta_h\Omega_h)}{1 + \tanh^2(r)\exp(\beta_h \Omega_h)}\Bigg],
\end{equation}
which coincides with the value found in Refs.~\cite{HuangPRE2012,CorreaSR2014,AlickiNJP2015,LatuneSR2019}.

\begin{figure}[t]
\begin{centering}
\includegraphics[width=1\columnwidth]{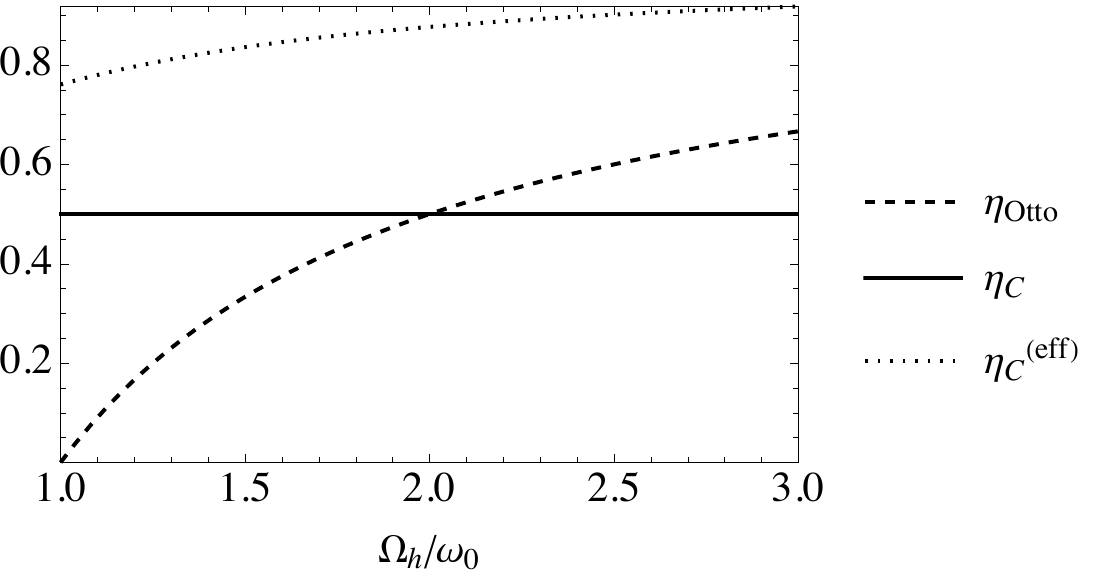}
\caption{Plot of Otto, Carnot and effective Carnot efficiency, for squeezing parameter $r = 0.3, \; \beta_c = 10/\omega_0, \; \beta_h = 5/\omega_0, \; \Omega_c = \omega_0$.
}
\label{fig:effectiveeff}
\end{centering}
\end{figure}

The fact that the squeezed reservoir has a higher effective temperature modifies the ranges of the operating modes as a function of the frequencies illustrated in Fig.~\ref{fig:thermalmachines}. 
Let us start by looking at the case of an engine. The condition for this kind of functioning now becomes
\begin{equation}
\label{eq:engcond}
\beta^{(\text{eff})}_h \Omega_h < \beta_c \Omega_c \quad \textrm{and}\quad \Omega_c < \Omega_h,
\end{equation}
while the average efficiency for slow driving is still the Otto efficiency
\begin{equation}
 {\overline{\eta^{(2)}}}  = 1 - \frac{\Omega_c}{\Omega_h} = \eta_{\rm Otto}.
\end{equation}
Combining the conditions given in Eq.~\eqref{eq:engcond} we get
\begin{equation}
\frac{\beta_h^{(\text{eff})}}{\beta_c} < \frac{\Omega_c}{\Omega_h} < 1.    
\end{equation}
By definition $\beta^{(\text{eff})}_h < \beta_h$, so squeezing decreases the lower bound on $\Omega_c / \Omega_h$, therefore allowing a larger average efficiency  {$ \overline{\eta^{(2)}}$} to be attained. This, as also discussed in Refs.~\cite{SinghPRE2020} and \cite{AbahEPL2014,CorreaSR2014,AlickiNJP2015,ManzanoPRE2016,KlaersPRX2017,ManzanoPRE2018,LatuneSR2019,HuangPRE2012}, is not in contradiction with the Carnot bound which is derived assuming equilibrium baths. Moreover, it does not violate the second law of thermodynamics provided that one accounts for the preparation of squeezed ancillas with additional resources. 

In this spirit, it is fairer to define a modified Carnot efficiency as
\begin{equation}
\eta_{C}^{(\text{eff})} = 1- \frac{\beta_h^{(\text{eff})}}{\beta_c},
\end{equation}
such that  {$\overline{\eta^{(2)}} \leq \eta_{C}^{(\text{eff})}$}.
Figure~\ref{fig:effectiveeff} plots the Otto efficiency obtained for the squeezed reservoir as well as the Carnot, and the effective Carnot efficiency, as a function of $\Omega_h$ for a squeezing parameter $r = 0.3$. As expected, the efficiency $\eta_{\rm Otto}$ may become larger than $\eta_C$ but never larger than $\eta_{C}^{(\text{eff})}$.

Now, let us move to the case of a refrigerator. The condition for this kind of functioning now becomes
\begin{equation}
\label{eq:effcond}
\beta_c \Omega_c < \beta^{(\text{eff})}_h \Omega_h \quad \textrm{and}\quad \Omega_c < \Omega_h,
\end{equation}
while the coefficient of performance for slow driving is still given by Eq.~\eqref{eqeff1}.
In order to have the largest possible COP, we want $\Omega_c / \Omega_h$ to be as close to $1$ as possible. However, combining the conditions in Eq.~\eqref{eq:effcond} gives
\begin{equation}
\frac{\Omega_c}{\Omega_h} < \frac{\beta_h^{(\text{eff})}}{\beta_c} < 1  , 
\end{equation}
meaning the upper bound on $\Omega_c / \Omega_h$ is decreased by squeezing and therefore so is the COP. Summarizing, while squeezing improves the performance of the two-oscillator system operating as an engine, it degrades it when operating as a refrigerator.

The squeezed bath also allows the creation of steady-state entanglement between the two oscillators that would otherwise be impossible to find in our model with thermal baths.  {The absence of entanglement in the steady state for thermal baths can be analytically proven in the slow driving limit under symmetry conditions, e.g. equal system's frequencies and equal couplings to the respective baths (see Appendix~\ref{sec:entanglement}). We have also verified numerically that there is no entanglement when these conditions are relaxed. This conclusion refers to our specific model and does not preclude the existence of entanglement in the steady state of quantum harmonic oscillators described by a different model.} 

\begin{figure}[t]
\begin{center}
\includegraphics[width=0.7\columnwidth]{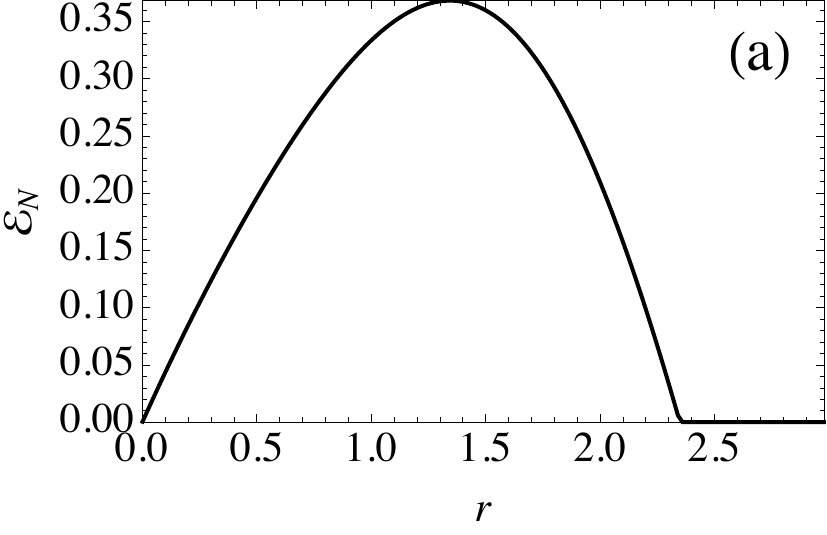}\\
\includegraphics[width=0.7\columnwidth]{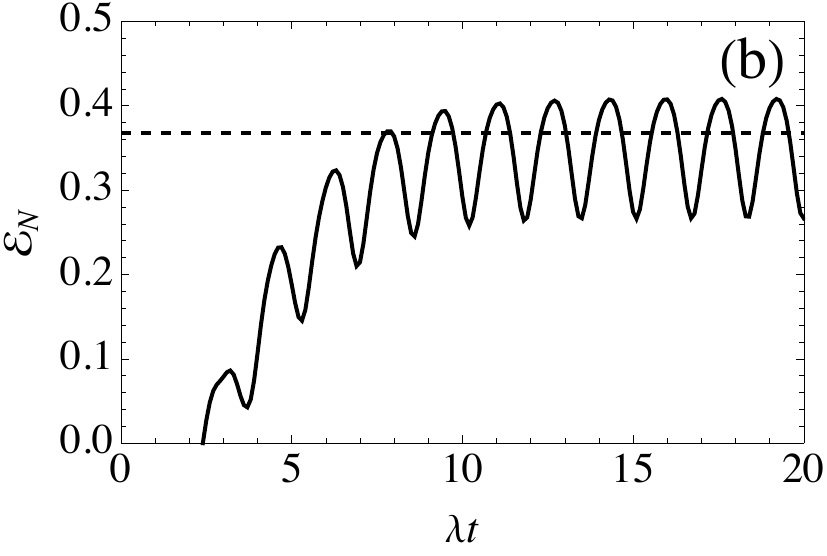}
\caption{(a) Steady-state logarithmic negativity $\mathcal{E}_N$ of the two system's oscillators with constant frequencies as a function of the squeezing parameter $r$ of the hot environment. 
(b) Logarithmic negativity $\mathcal{E}_N$ (solid line) as a function of time for  $r = 1.3$ when we drive the two system oscillators. The dashed line is the maximum value attained with constant frequencies.
Parameters for both plots:  $\lambda = 1, \beta_c = 81.5/\lambda, \beta_h = 40.7/\lambda,  g_c = 0.495\sqrt \lambda, g_h = 0.612 \sqrt\lambda,\omega_{0,1}=\Omega_c = 4.07\lambda, \omega_{0,2}=\Omega_h = 0.244\lambda,
\delta\omega_{0,1} = \omega_{0,1} /10, \delta\omega_{0,2} = \omega_{0,2} /10, 
 \theta = \pi \lambda$.}
\label{fig:logneg}
\end{center}
\end{figure}

Figure~\ref{fig:logneg} shows an example of this  {squeezing-induced} entanglement as measured by the logarithmic negativity (see Appendix~\ref{sec:entanglement}) for oscillators with constant frequencies (no modulation, i.e., $\theta=0)$. Such entanglement is quite fragile and requires the baths to be at very low temperatures. 
Notice that the value of the logarithmic negativity increases from zero as we introduce squeezing until a maximum is reached. From this point, increasing the squeezing parameter, $r$, only decreases the logarithmic negativity until it disappears for larger $r$.
Choosing the value of $r$ which maximizes the logarithmic negativity, Fig.~\ref{fig:logneg} also shows that fast driving the frequency of the oscillators can lead, with appropriately optimized parameters, to higher instantaneous values of the logarithmic negativity compared to the slow driving case with constant frequencies.

Summing up: for the thermodynamic quantities, the presence of squeezing only raises the effective temperature of the squeezed bath and does not change qualitatively the overall behavior; for the quantum correlations, instead, squeezing has a pivotal role in the creation of entanglement.

\section{Conclusions}
\label{sec:conclusions}
We have investigated the dissipative evolution of a network of driven QHO's and studied the energy flows in and out of the system in terms of heat currents and  {power}. A system composed of both one and two driven QHO's coupled to two heat baths of different temperatures may operate as an engine, a refrigerator, or an accelerator when the system is parametrically driven slowly. 
The efficiency of such an engine oscillates periodically in time around the Otto efficiency. While the instantaneous coefficient of performance of a refrigerator also oscillates, its average over a cycle can be found to be above the corresponding value of an Otto refrigerator. A higher maximum value and average over a cycle of the instantaneous COP has been found when we have added a second driven oscillator to the system.

As we have increased the driving speed in both the one- and two-oscillator systems, we have found that there is an increase in the  {power} injected into the system and it has been found that there is some resonant driving frequency for which a maximum is reached.

We have also shown that applying squeezing to the hot bath effectively raises its temperature and leads to a larger effective Carnot efficiency.
We have  found that squeezing is a necessary requirement for the sustained generation of entanglement within a system of two driven dissipative QHO's. Controlled driving of the system's oscillators can increase the degree of entanglement.

Our findings can be readily verified in experimental platforms described by linearly coupled quantum harmonic oscillators, e.g., trapped ions, atomic ensembles, or photonic systems. The extension of our model to a large number of oscillators and the corresponding scaling of thermodynamic quantities will be the subject of further studies. 

\acknowledgements
G.D.C. acknowledges support by the UK EPSRC EP/S02994X/1. H.L. acknowledges
support from the UK EPSRC 2278075. N.P. acknowledges the “Universit\'e de Franche-
Comt\'e” for financial support throughout the mobility grant “MOBILIT\'E
INTERNATIONALE DES DOCTORANTS 2019” for his visit at the Queen’s University
Belfast (QUB) and thanks the QTeQ group and CTAMOP at QUB for their kind
hospitality on this occasion. B.B. acknowledges support by the French
“Investissements d’Avenir” program, Project ISITE-BFC (Contract No. ANR-15-IDEX-
03).
  
  \section*{Author Declarations}
\subsection*{Conflict of Interest}
The authors have no conflicts to disclose.

\appendix
\section{Expression of the  {power} for two driven oscillators}
\label{sec:appa}
The expression for the work done on two oscillators in contact with two baths is
\begin{eqnarray}
\dot{W}^{(2)} & =& \frac{\lambda \left( \omega_1 - \omega_2 \right)}{\sqrt{ \omega_1 \omega_2}}\left( \omega_1  \sigma_{x_1p_2 } - \omega_2 \sigma_{x_2p_1}  \right) 
\nonumber\\   
&+& \frac{g_c^2}{2} \left[ \omega_1  \left( 1 + 2 n_c - \omega_1  \sigma_{x_1x_1}  \right) - \sigma_{p_1p_1}  \right]
\nonumber\\   
& + &\frac{g_h^2}{2} \left[ \omega_2  \left( 1 + 2 n_h - \omega_2  \sigma_{x_2x_2}  \right) - \sigma_{p_2p_2}  \right]
\nonumber\\   
&+& \omega_1  \dot{\omega}_1  \sigma_{x_1x_1} 
+ \omega_2  \dot{\omega}_2  \sigma_{x_2x_2} 
\nonumber\\   
&-& \dot{Q}_c^{(2)}  - \dot{Q}_h^{(2)},     
\end{eqnarray}
where for simplicity we have removed the explicit time-dependence of various quantities.
In the slow driving limit ($\dot\sigma_S=0$), the above expression becomes
\begin{eqnarray}
\dot{W}^{(2)} & =& \frac{(2n_c+1)\dot\omega_1+(2n_h+1)\dot\omega_2}{2}+
\nonumber \\
&+&\frac{4\lambda^2(g_c^2+g_h^2)(\dot\omega_1 g_c^2-\dot\omega_2 g_h^2)(n_1-n_2)}{\left(g_c^2 + g_h^2\right)^2 \left(4 \lambda^2 + g_c^2 g_h^2\right) + 4 g_c^2 g_h^2 [\omega_1 - \omega_2]^2}\nonumber
\\
& -& \dot{Q}_c^{(2)} - \dot{Q}_h^{(2)}.
\end{eqnarray}
The expression for the  {power} under the assumption of $\omega_1(t)=\omega_2(t)=\omega(t)$ simplifies to

\begin{widetext}
\begin{eqnarray}
\dot{W}^{(2)}  = \frac{\left\{\frac{g_c^2 g_h^2}{\lambda^2} \left(g_c^2 + g_h^2\right) \left(1 + n_c + n_h\right) +  4\left[ g_c^2 (1 + 2 n_c) + g_h^2 (1 + 2 n_h)\right]\right\} \dot{\omega}
- 4 g_c^2 g_h^2 \left(n_c - n_h\right) \left(\Omega_c - \Omega_h\right)}{\left(g_c^2 + g_h^2\right) \left(4 + \frac{g_c^2 g_h^2}{\lambda^2}\right)}.
\end{eqnarray}
\end{widetext}

\section{Entanglement  of Gaussian states}
\label{sec:entanglement}
In this section, we review the logarithmic negativity as an entanglement quantifier for two quantum harmonic oscillators ~\cite{VidalPRA2002,PlenioPRL2005}. We follow the method detailed, for instance, in Ref.~\cite{AdessoJPA2007}, and redefine our position and momentum operators as
\begin{equation}
\tilde{x}_\alpha = \sqrt 2 x_\alpha, \;\;\;\;\; \tilde{p}_\alpha = \sqrt 2 p_\alpha,
\end{equation}
where $\alpha = 1, \;2$. 
We define the vector of the system's quadratures
\begin{equation}
\tilde{R} = \Big\{\tilde{x}_1,\; \tilde{p}_1,\;\tilde{x}_2,\; \tilde{p}_2  \Big\},    
\end{equation}
the matrix
\begin{equation}
\tilde{A}_S = \frac{\omega(t)}{4} \mathbb{I}_4 ,
\end{equation}
proportional to the $4\times 4$ identity matrix $\mathbb{I}_4$
and the system's covariance matrix
\begin{equation}
\tilde{\sigma}_{ij} = \frac{1}{2}\expval{\tilde{R}_i \tilde{R}_j + \tilde{R}_j \tilde{R}_i} - \langle \tilde{R}_i\rangle \langle \tilde{R}_j \rangle.
\end{equation}
Let us rewrite the system's covariance matrix as
\begin{equation}
\tilde{\sigma}_S(t) = \begin{pmatrix}
\boldsymbol{A} & \boldsymbol{C}\\
\boldsymbol{C}^T & \boldsymbol{B}
\end{pmatrix}    ,
\end{equation}
where
\begin{equation}
\boldsymbol{A} = \begin{pmatrix}
\tilde{\sigma}_{\tilde{x}_1\tilde{x}_1}(t) & \tilde{\sigma}_{\tilde{x}_1\tilde{p}_1}(t)\\
\tilde{\sigma}_{\tilde{x}_1\tilde{p}_1}(t) & \tilde{\sigma}_{\tilde{p}_1\tilde{p}_1}(t)
\end{pmatrix}  ,
\end{equation}
\begin{equation}
\boldsymbol{B} = \begin{pmatrix}
\tilde{\sigma}_{\tilde{x}_2\tilde{x}_2}(t) & \tilde{\sigma}_{\tilde{x}_2\tilde{p}_2}(t)\\
\tilde{\sigma}_{\tilde{x}_2\tilde{p}_2}(t) & \tilde{\sigma}_{\tilde{p}_2\tilde{p}_2}(t)
\end{pmatrix}  ,
\end{equation}
\\
\begin{equation}
\boldsymbol{C} = \begin{pmatrix}
\tilde{\sigma}_{\tilde{x}_1\tilde{x}_2}(t) & \tilde{\sigma}_{\tilde{x}_1\tilde{p}_2}(t)\\
\tilde{\sigma}_{\tilde{x}_2\tilde{p}_1}(t) & \tilde{\sigma}_{\tilde{p}_1{p}_2}(t)
\end{pmatrix}.
\end{equation}
We now define 
\begin{eqnarray}
I_1 = \det[\boldsymbol{A}],  \;\;\;\;\; & I_2 = \det[\boldsymbol{B}], \nonumber \\
I_3 = \det[\boldsymbol{C}],  \;\;\;\;\; & I_4 = \det[\tilde{\sigma}_S(t)].
\end{eqnarray}
In order to compute the logarithmic negativity, we need to calculate the symplectic eigenvalues of the partially-transposed covariance matrix which, after some algebra, read
\begin{equation}
\tilde\nu_{\pm} = \sqrt{\frac{1}{2}\left( \Tilde{\Lambda}  \pm \sqrt{\Tilde{\Lambda} ^2 - 4 I_4} \right)},
\end{equation}
where
\begin{equation}
\Tilde{\Lambda} = I_1 + I_2 - 2I_3.
\end{equation}
The logarithmic negativity is then defined in terms of the smallest symplectic eigenvalue $\tilde\nu_-$ as
\begin{equation}
\mathcal{E}_N=\max(0,-\ln\tilde{\nu}_-).
\end{equation}
 {The expression of the logarithmic negativity is generally complex but can be simplified in special cases. For instance, applying the limiting cycle solutions, without squeezing and with the additional conditions $g_c = g_h = g$ and $\omega_1(t) = \omega_2(t) = \omega(t)$, we get
\begin{widetext}
\begin{eqnarray}
\Tilde{\nu}_- &=& \frac{1}{g^4+4\lambda^2}\Bigg[g^8 \Big( 1 + 2 n_c (1+n_c)+ 2 n_h (1+n_h) \Big) +4g^4\lambda^2\Big(2+n_c(4+n_c)+n_h(4+n_h) + 6n_c n_h)\Big) +16\lambda^4(1+n_c+n_h)^2 \nonumber\\
&-& 2\Bigg[g^8(n_c - n_h)^2 \Big(g^8 (1 + n_c + n_h)^2
+ 4 g^4 \lambda^2(2 + n_c(4+n_c)+n_h(4+n_h) + 6n_c n_h) + 
   16 \lambda^4(1 + n_c + n_h)^2 \Big)  \Bigg]^{1/2}\Bigg]^{1/2}.\nonumber\\
\end{eqnarray}
\end{widetext}
For entanglement to be present, we must have $\Tilde{\nu}_- < 1$. We find that 
\begin{equation}
\lim_{\lambda\rightarrow \pm \infty} \Tilde{\nu}_- = 1 + n_c +n_h   \ge 1,
\end{equation}
as $n_c,n_h \ge 0$. The values of $\lambda$ that give $\Tilde{\nu}_- = 1$ are
\begin{eqnarray}
\lambda &=& \pm \ii g^2 \sqrt{\frac{n_c(1+n_h)}{(n_c + n_h)(2+n_c +n_h)}}, \nonumber\\
\lambda &=& \pm \ii g^2\sqrt{\frac{n_h(1+n_c)}{(n_c + n_h)(2+n_c +n_h)}}.
\end{eqnarray}
Since the coupling constant $\lambda \in \mathbb{R}$, we cannot have entanglement with these conditions as there is no $\lambda$ for which $\Tilde{\nu}_- < 1$. }

  \section*{Data availability}
The data that support the findings of this study are available from the corresponding author upon reasonable request.

\section*{References}
\bibliography{bib}
\end{document}